\documentclass[12pt]{JHEP3}
\usepackage{epsfig,amsmath,amssymb}

\newcommand{\lsim}{\mathrel{\mathop{\kern 0pt \rlap
  {\raise.2ex\hbox{$<$}}}
  \lower.9ex\hbox{\kern-.190em $\sim$}}}
\newcommand{\gsim}{\mathrel{\mathop{\kern 0pt \rlap
  {\raise.2ex\hbox{$>$}}}
  \lower.9ex\hbox{\kern-.190em $\sim$}}}

\preprint{KIAS-P010015}

\title{Dark matter and a new gauge boson through kinetic mixing}

\author{Eung Jin Chun  and Jong-Chul Park \\
\\Korea Institute for Advanced Study\\
Heogiro 87, Dongdaemun-gu\\
Seoul 130-722, Korea\\
Emails: \email{ejchun@kias.re.kr, jcpark@kias.re.kr} }
\author{Stefano Scopel \\
Department of Physics, Sogang University\\
1 Sinsu-dong, Mapo-gu\\
        Seoul 121-742, Korea\\
Email: \email{scopel@sogang.ac.kr} }

\abstract{ We consider a hidden sector model of dark matter which is
  charged under a hidden $U(1)_X$ gauge symmetry. Kinetic mixing of
  $U(1)_X$ with the Standard Model hypercharge $U(1)_Y$ is allowed to
  provide communication between the hidden sector and the Standard
  Model sector. We present various limits on the kinetic mixing
  parameter and the hidden gauge coupling constant coming from various
  low energy observables, electroweak precision tests, and the right
  thermal relic density of the dark matter.  Saturating these
  constraints, we show that the spin-independent elastic cross section
  of the dark matter off nucleons is mostly below the current
  experimental limits, but within the future sensitivity. Finally, we
  analyze the prospect of observing the hidden gauge boson through its
  dimuon decay channel at hadron colliders.}


\maketitle

\begin{document}

\section{Introduction}

One of the popular scenarios for a TeV-scale physics beyond the
Standard Model is postulating an additional gauge interaction other
than the Standard Model one $SU(3)_c \times SU(2)_L \times U(1)_Y$. A
classic example is an extra $U(1)$ interaction that arises from a
grand unification theory \cite{Langacker08}. Such a possibility has
been well studied as it might be discovered in the early stage of the
LHC experiment. Another interesting possibility for an extended gauge
sector is to assume a $U(1)_X$ interaction in the hidden sector, in
the sense that Standard Model particles are neutral under
$U(1)_X$. However, the hidden sector and the Standard Model sector can
couple to each other through the kinetic mixing of $U(1)_X$ and
$U(1)_Y$ \cite{Holdom85}. Since the kinetic mixing term is
gauge-invariant, it can be present at the tree-level.  In the hidden
$U(1)_X$ model, one can introduce a massive Dirac fermion charged
under $U(1)_X$ which is a dark matter (DM) candidate.  This kind of
scenario has been used to implement MeV DM \cite{Huh07}, a sub-GeV $X$
boson for the Sommerfeld enhancement employing kinetic and Higgs
mixing \cite{Chun08}, light DM with Sommerfeld enhancement in the
NMSSM with gauge mediation \cite{Kang10}, a TeV scale hidden sector
through Higgs mixing \cite{Gopalakrishna09}, and a 10 GeV DM through
kinetic mixing \cite{Mambrini10}.  Let us also remark that Stuekelberg
$Z'$ models have similar features~\cite{Cheung07,Feldman07,Fucito08}.

In this paper, we examine various phenomenological implications of
the kinetic mixing of $U(1)_X$ and of a DM candidate which has a
preferable mass scale of $0.1 - 1$ TeV. In this model, the hidden
dark matter physics is described by four parameters: the kinetic
mixing angle $\epsilon$, the dark matter mass $m_\psi$, the $X$
boson mass $m_X$, and the $X$ gauge coupling $g_X$. These
parameters are constrained by various low-energy and electroweak
observables, Tevatron II results, and the thermal relic density of
dark matter. After considering these constraints, we will look for
perspectives for the direct detection of dark matter and the LHC
discovery of the $X$ boson.

This paper is organized as follows. In Section \ref{interactions},
we set up the hidden sector dark matter model and present
interaction vertices relevant for further discussions.  In Section
\ref{sec:low_energy_constraints}, we work out various constraints
on the kinetic mixing parameter $\epsilon$ from low-energy
observables \cite{Bouchiat04,Fayet07} and electroweak precision
tests \cite{Kumar06,Chang06} used in the phenomenological
discussion of the ensuing Sections. In Section \ref{abundance},
the DM annihilation rate is calculated and normalized to the
observed DM relic density. In particular, by using this procedure
we fix a combination of $\epsilon$ and $g_X$ ($\simeq \epsilon
g_X$ in the limit $\epsilon \ll 1$) as a function of the two
masses $m_\psi$ and $m_X$. In Section \ref{sec:direct}, the
spin-independent DM-nucleon cross section is calculated and
compared to the recent CDMS II~\cite{CDMS2} and
XENON100~\cite{XENON100} results. In Section \ref{sec:tevatron},
we analyze the Tevatron II limit on the $X$ boson production and
dimuon decay and the LHC perspective for detection of the same
quantity, which depends on the branching ratio of the X boson
decay to the dark matter particle. Finally, we conclude in Section
\ref{sec:conclusions}.

\section{Hidden $U(1)_X$ model and gauge interactions} \label{interactions}

We consider a hidden sector containing a gauge symmetry $U(1)_X$ and a
Dirac fermion dark matter candidate at the TeV scale, which couples to
the Standard Model sector through kinetic mixing.  The full Lagrangian
including kinetic mixing is:
\begin{equation}
{\cal L}= {\cal L}_{SM} -{1\over2} \sin\epsilon\, \hat{B}_{\mu\nu} \hat{X}^{\mu\nu} -\frac{1}{4}\hat{X}^{\mu\nu}\hat{X}_{\mu\nu}
 - g_X \hat{X}^\mu \bar{\psi}\gamma_\mu \psi + {1\over2} m_{\hat{X}}^2 \hat{X}^2
  + m_\psi \bar{\psi}\psi,
\end{equation}
where the $U(1)_X$ is assumed to be broken spontaneously leading to the gauge boson mass $m_{\hat{X}}$.
In the Standard Model sector, the $\hat{Z}$ gauge boson has the mass $m_{\hat{Z}}$ and the
gauge couplings are denoted by $\hat{g}=\hat{e}/s_{\hat{W}}$ and
$\hat{g}'= \hat{e}/c_{\hat{W}}$.  Diagonalizing away the kinetic
mixing term and mass mixing terms is made by the following
transformation \cite{babu97}:
\begin{eqnarray} \label{transformation}
 \hat{B} &=& c_{\hat{W}} A - (t_\epsilon s_\xi+ s_{\hat{W}} c_\xi) Z
 + (s_{\hat{W}} s_\xi-t_\epsilon c_\xi) X \,, \nonumber\\
 \hat{W}_3 &=& s_{\hat{W}} A + c_{\hat{W}} c_\xi Z
 - c_{\hat{W}} s_\xi X \,, \nonumber\\
 \hat{X} &=& {s_\xi \over c_\epsilon} Z + {c_\xi\over c_\epsilon} X \,,
\end{eqnarray}
where the angle $\xi$ is determined by:
\begin{equation}
 \tan2\xi = - {m_{\hat{Z}}^2 s_{\hat{W}} \sin2\epsilon \over
               m_{\hat{X}}^2 - m_{\hat{Z}}^2
               (c^2_\epsilon-s^2_\epsilon s_{\hat{W}}^2) } \,.
\label{eq:t_csi}
\end{equation}
Then, the $X$ and $Z$ gauge bosons get the redefined masses:
\begin{eqnarray}
 m_Z^2 &=& m_{\hat{Z}}^2(1+s_{\hat{W}} t_\xi t_\epsilon) \,,  \label{eq:mzhat} \\
 m_X^2 &=&
   { m_{\hat{X}}^2 \over c_\epsilon^2 (1+s_{\hat{W}} t_\xi t_\epsilon)} \,.
\label{eq:mx}
\end{eqnarray}
Moreover, from Eqs.~(\ref{eq:t_csi}) - (\ref{eq:mx}) one can find
$t_\xi$ as a function of $r_X\equiv m_X^2/m_Z^2$:
\begin{eqnarray}
t_\xi &=& - \frac{1}{s_{\hat{W}} t_\epsilon},\label{eq:t_csi1}\\
t_\xi&=& \frac{1-r_X \pm \sqrt{(1-r_X)^2-4s_{\hat{W}}^2
t_\epsilon^2 r_X}}{2s_{\hat{W}} t_\epsilon r_X}\;. \label{eq:t_csi2}
\end{eqnarray}
The solution given by Eq.(\ref{eq:t_csi1}) is unphysical since it
corresponds to $m_Z = 0$ and $m_X = \infty$.  In addition, the other
two solutions (\ref{eq:t_csi2}) are physical only in the range $r_X
\leq 1 + 2s_{\hat{W}}^2 t_\epsilon^2 - 2\sqrt{s_{\hat{W}}^2
  t_\epsilon^2(1+s_{\hat{W}}^2 t_\epsilon^2)}$ or $r_X \geq 1 +
2s_{\hat{W}}^2 t_\epsilon^2 + 2\sqrt{s_{\hat{W}}^2
  t_\epsilon^2(1+s_{\hat{W}}^2 t_\epsilon^2)}$, where
$(1-r_X)^2-4s_{\hat{W}}^2 t_\epsilon^2 r_X^2 > 0$. In the following
analysis we exclude the unphysical region, which is limited to a small
interval around $m_X=m_Z$.  As we will see shortly, the weak mixing
angle $s_{\hat{W}}$ is very close to the physical angle $s_W$ due to
the strong $\rho$ parameter limit.

Let us now list all the interaction vertices relevant for our analysis.
The $W$, $Z$, and $X$ gauge bosons have the following fermion
couplings:
\begin{eqnarray}
 {\cal L}_W &=& -{e\over \sqrt{2} s_{\hat{W}} }\, W^+_\mu\,
      \left\{ \bar{\nu} \gamma^\mu P_L e + \bar{u} \gamma^\mu P_L d \right\}
       + c.c.\,, \label{Wfermion} \\
 {\cal L}_Z &=& -{e\over c_{\hat{W}} s_{\hat{W}} }\, c_\xi \,
    Z_\mu\, \bar{f} \gamma^\mu \left\{
         P_L \left[ T_3 ( 1 + s_{\hat{W}} t_\epsilon t_\xi )
          - Q (s_{\hat{W}}^2 + s_{\hat{W}} t_\epsilon t_\xi ) \right] \right. \nonumber\\
  &&~~~~~~~~~~~~~~~~~~~~~~~\left. - P_R \left[Q (s_{\hat{W}}^2
      + s_{\hat{W}} t_\epsilon t_\xi)\right] \right\} f
      \, - \, g_X{s_\xi\over c_\epsilon} Z_\mu\,\bar{\psi}\gamma^\mu\psi \,, \label{Zfermion} \\
 {\cal L}_X &=& -{e\over c_{\hat{W}} s_{\hat{W}} }\, c_\xi\,
     X_\mu\, \bar{f} \gamma^\mu \left\{
         P_L \left[ T_3 (s_{\hat{W}} t_\epsilon-t_\xi)
                    + Q (s_{\hat{W}}^2 t_\xi - s_{\hat{W}} t_\epsilon) \right]
         \right.\nonumber\\
  &&~~~~~~~~~~~~~~~~~~~~~~~\left. + P_R \left[Q (s_{\hat{W}}^2 t_\xi
                                     - s_{\hat{W}} t_\epsilon)\right] \right\} f
       \,-\, g_X{c_\xi\over c_\epsilon} X_\mu\,\bar{\psi}\gamma^\mu\psi \,.\label{Xfermion}
\end{eqnarray}
The couplings of the $Z$ and $X$ gauge bosons to the $W$ gauge
bosons are given by:
\begin{eqnarray} \label{VWW}
 {\cal L}_{VWW} = {e \over t_{\hat{W}}}\, c_\xi\, [[Z W^+ W^-]]
              - {e \over t_{\hat{W}}}\, s_\xi\, [[X W^+ W^-]]\,,
\end{eqnarray}
where $[[V W^+ W^-]] \equiv i [(\partial_\mu W_\nu^+ -
\partial_\nu W_\mu^+) W^{\mu-} V^\nu - (\partial_\mu W_\nu^- -
\partial_\nu W_\mu^-) W^{\mu+} V^\nu + (1/2) (\partial_\mu V_\nu -
\partial_\nu V_\mu) (W^{\mu+}W^{\nu-} - W^{\mu-}W^{\nu+})]$.
The couplings of the Higgs scalar $h$ to the $Z$ and $X$ gauge
bosons are:
\begin{eqnarray} \label{hVV}
 {\cal L}_{hVV} &=& {m_{\hat{Z}}^2\over v}\, c_\xi^2\, h
   \left[ (1 + s_{\hat{W}}t_\xi t_\epsilon)^2 Z_\mu Z^\mu
     + t_\xi^2 \left(1 - s_{\hat{W}}{t_\epsilon\over t_\xi} \right)^2 X_\mu X^\mu
  \right. \nonumber\\
  &&~~~~~~~~~~~\left. + 2 t_\xi \left(-1 + 2s_{\hat{W}}{t_\epsilon\over t_\xi}
                         + s_{\hat{W}}^2t_\epsilon^2 \right) X_\mu Z^\mu \right]\,.
\end{eqnarray}
Summarizing the interaction vertices in Eqs.~(\ref{Wfermion}) - (\ref{hVV}), let us
define the various couplings, $g$'s, as follows:
\begin{eqnarray}\label{g's}
 {\cal L} &=& W_\mu^+\, g_f^W [ \bar{\nu} \gamma^\mu P_L e
                        + \bar{u} \gamma^\mu P_L d ] + c.c. \nonumber \\
  &+& Z_\mu \left[ g^Z_{fL}\, \bar{f} \gamma^\mu P_L f
                          + g^Z_{fR}\, \bar{f} \gamma^\mu P_R f
       + g^Z_\psi\, \bar{\psi}\gamma^\mu\psi \right] + g_W^Z [[Z W^+ W^-]] \nonumber \\
  &+& X_\mu \left[ g^X_{fL}\, \bar{f} \gamma^\mu P_L f
                + g^X_{fR}\, \bar{f} \gamma^\mu P_R f
       + g^X_\psi\, \bar{\psi}\gamma^\mu\psi \right] + + g_W^X [[X W^+ W^-]] \nonumber \\
  &+& h \left[  g^h_{ZZ}\, Z_\mu Z^\mu + g^h_{XX} X_\mu X^\mu + g^h_{XZ} X_\mu Z^\mu
       \right] \,.
\end{eqnarray}

Unlike the $Z$ and $X$ gauge bosons, the mass of the $W$ gauge
boson is not modified by the above
transformation~(\ref{transformation}):
\begin{equation}
m_W^2 = m_{\hat{W}}^2 = m_{\hat{Z}}^2c_{\hat{W}}^2 \,.
\end{equation}
Then, the $\rho$ parameter is given by:
\begin{equation}
\rho \equiv {m_W^2\over m_Z^2 c_W^2} = {c_{\hat{W}}^2\over (1 +
s_{\hat{W}} t_\xi t_\epsilon) c_W^2}\,.
\end{equation}
Consequently, the current bound on the $\rho$ parameter, $\rho-1 =
4^{+8}_{-4} \times 10^{-4}$~\cite{PDG}, provides a constraint on the
parameter $\epsilon$ as a function of the gauge boson masses.  Let us
note that the photon coupling does not change ($\hat{e}=e$) and the
identity $m_Z^2/(g^2+g'^2)=m_{\hat{Z}}^2/(\hat{g}^2+\hat{g}'^2)$
\cite{babu97} leads to the relation between the original and redefined
weak mixing angles:
\begin{equation}
 c_W^2 s_W^2= {c_{\hat{W}}^2 s_{\hat{W}}^2 \over 1+ s_{\hat{W}}
 t_\xi t_\epsilon}\,.
\end{equation}
Therefore, the $\rho$ parameter can be recast as
\begin{equation}
\rho = {s_W^2 \over s_{\hat{W}}^2 }\,.
\end{equation}
Defining $\hat{\delta} \equiv \rho -1 = {s_W^2 / s_{\hat{W}}^2 }-1$
and expanding at the leading order in $\hat{\delta}$, we find:
\begin{equation} \label{delta}
 \omega\equiv  s_W t_\xi t_\epsilon \simeq -(1 - t_W^2 ) \hat{\delta}\,.
\end{equation}
%Here, $\hat{\delta}$ is determined as follows:
%\begin{eqnarray} \label{delta}
% \hat{\delta} &\simeq&
% { 2 s_W^2 c_W^2 t_\epsilon^2 ( r_X -1 - s_W^2 t_\epsilon^2 ) \over
%  (r_X^2 + 1) (2-4 s_W^2) - r_X (4 - 8 s_W^2 -s_W^2 c_W^2 t^2_\epsilon )
%   - s_W^2 t^2_\epsilon (5 - 9 s_W^2) + s_W^4 t^4_\epsilon (1-3s_W^2) } \,,~~~~~~~~
%\end{eqnarray}
%where $r_X\equiv m_X^2/m_Z^2$.

In the redefined physical basis, the above couplings can be
rewritten in the first order of $\omega$ (or $\hat{\delta}$) as
follows:
\begin{eqnarray}
 {\cal L}_W &=& -{e\over \sqrt{2} s_W}
 \left(1-{\omega \over 2(1-t_W^2)}\right)\, W^+_\mu\,
 \left\{ \bar{\nu} \gamma^\mu P_L e + \bar{u} \gamma^\mu P_L d \right\} + c.c.\,, \label{interaction-Wff}\\
 {\cal L}_Z &=& -{e\over c_{{W}} s_{{W}} }\, c_\xi \,
    Z_\mu\, \bar{f} \gamma^\mu
    \left\{ P_L T_3 \left[1+ {\omega \over 2}\right]
  - Q \left[s_{{W}}^2 + \omega \left( {2 - t_W^2 \over 2( 1- t_W^2) }\right) \right] \right\} f \nonumber\\
  &&
      -\, g_X {s_\xi\over c_\epsilon}\, Z_\mu\,\bar{\psi}\gamma^\mu\psi \,, \label{interaction-Zff}\\
 {\cal L}_X &=& -{e\over c_{{W}} s_{{W}} }\, {c_\xi} \,
     X_\mu\, \bar{f} \gamma^\mu \left\{
         P_L T_3 \left[ s_W t_\epsilon - t_\xi + {1\over 2}\,
         \omega \left(t_\xi + { s_W t_W^2 t_\epsilon\over 1 - t_W^2} \right) \right]
         \right. \nonumber\\
  && ~~~~~~~~~~~~~~~~~~~~~~~~\left. + Q \left[ s_W^2 t_\xi - s_W t_\epsilon
         + {1\over 2}\, t_W^2 \omega \left({t_\xi -s_W t_\epsilon \over 1-t_W^2 } \right) \right]
         \right\} f \nonumber\\
  &&
       -\, g_X {c_\xi\over c_\epsilon} X_\mu\,\bar{\psi}\gamma^\mu\psi \,,  \label{interaction-Xff}\\
 {\cal L}_{VWW} &=& {e \over t_{{W}}}
     \left(1-{\omega \over 2( c_W^2-s_W^2)} \right)
               \left\{c_\xi\, [[Z W^+ W^-]]
              - s_\xi\, [[X W^+ W^-]] \right\}\,, \label{interaction-VWW}\\
 {\cal L}_{hVV} &=& {m_{{Z}}^2\over v}\, c_\xi^2\,  h
   \left\{ [1 + \omega] Z_\mu Z^\mu
       + \left[ t_\xi^2 + s_W^2 t_\epsilon^2
               - \omega \left(2 + t_\xi^2
               - { s_W^2 t_W^2 t_\epsilon^2 \over 1 - t_W^2} \right)
        \right] X_\mu X^\mu
  \right. \nonumber\\
&&   \left.
  + 2 \left[
       2s_W t_\epsilon - t_\xi + \omega \left( t_\xi + {s_W t_W^2 t_\epsilon
       \over 1- t_W^2} \right)
     \right] X_\mu Z^\mu
     \right\}\,. \label{interaction-hVV}
\end{eqnarray}
From these expressions, we can read off the couplings defined in
Eq.~(\ref{g's}). The explicit expressions of the redefined interaction couplings are given in
Appendix~\ref{couplings}.

\section{Low-energy and electroweak constraints}
\label{sec:low_energy_constraints}
The dark matter model with a hidden $U(1)_X$ sector introduced in
the previous Section has four free parameters: $\epsilon$, $g_X$,
$m_X$, and $m_\psi$. In this Section, we show how sizable the
$U(1)_X$ contributions are for the muon $g-2$, atomic
parity-violation, $\rho$ parameter and Electro--Weak Precision
Tests (EWPT). From these analyses, we will obtain an upper limit
on the kinetic mixing parameter $\epsilon$ as a function of the hidden
gauge boson mass $m_X$.

\subsection{$g_\mu-2$}

The exchange of the hidden gauge boson $X$ induces a contribution
to the anomalous magnetic moment of the muon, $a_\mu =
(g_\mu-2)/2$. The modified couplings of the $Z$ boson induce an
additional contribution. Adopting the formula in
Ref.~\cite{Fayet07}, we find:
\begin{eqnarray}
&& \delta a_\mu \approx
 { (g_{\mu V}^X)^2 - 5 \,(g_{\mu A}^X)^2
  \over 12\pi^2}{m_\mu^2 \over m_X^2}
 + \Delta { (g_{\mu V}^Z)^2 - 5\, (g_{\mu A}^Z)^2
  \over 12\pi^2}{m_\mu^2 \over m_Z^2}\,,
\end{eqnarray}
where $g_{\mu V,A}^{X,Z} \equiv g_{\mu R}^{X,Z} \pm g_{\mu
L}^{X,Z}$ are given in Appendix~\ref{couplings} and $\Delta$
represents the deviation from the value calculated in the SM due to
the modification of $Z$ couplings.
%
%
%%%%%%%%%%%%%%%%%%%%%%%%%%%%%%%%%%%%%%%%%%%%%%%%%%%%%%%%%%%%%%%
\begin{figure}
\begin{center}
\includegraphics[width=0.80\linewidth]{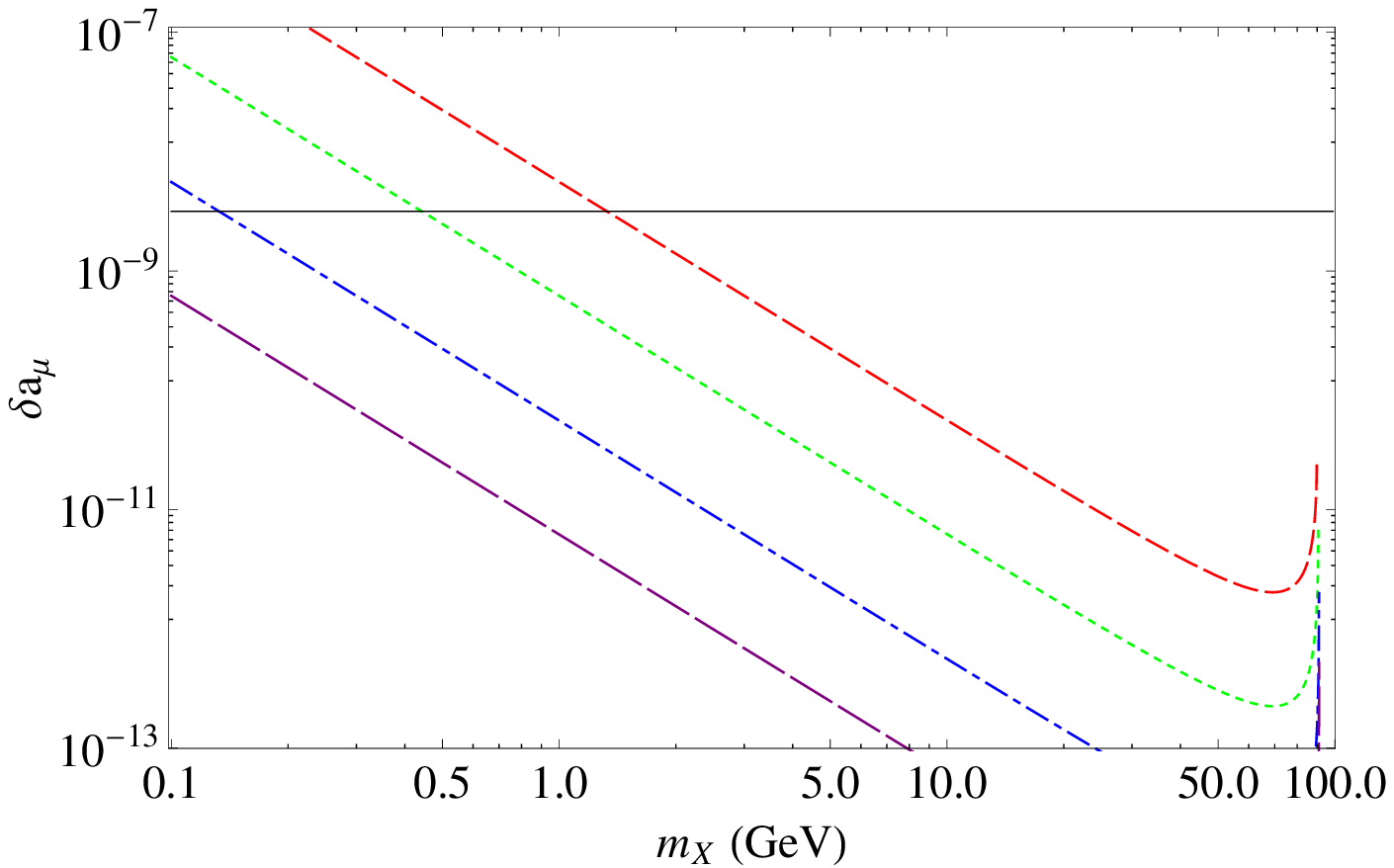}
\includegraphics[width=0.80\linewidth]{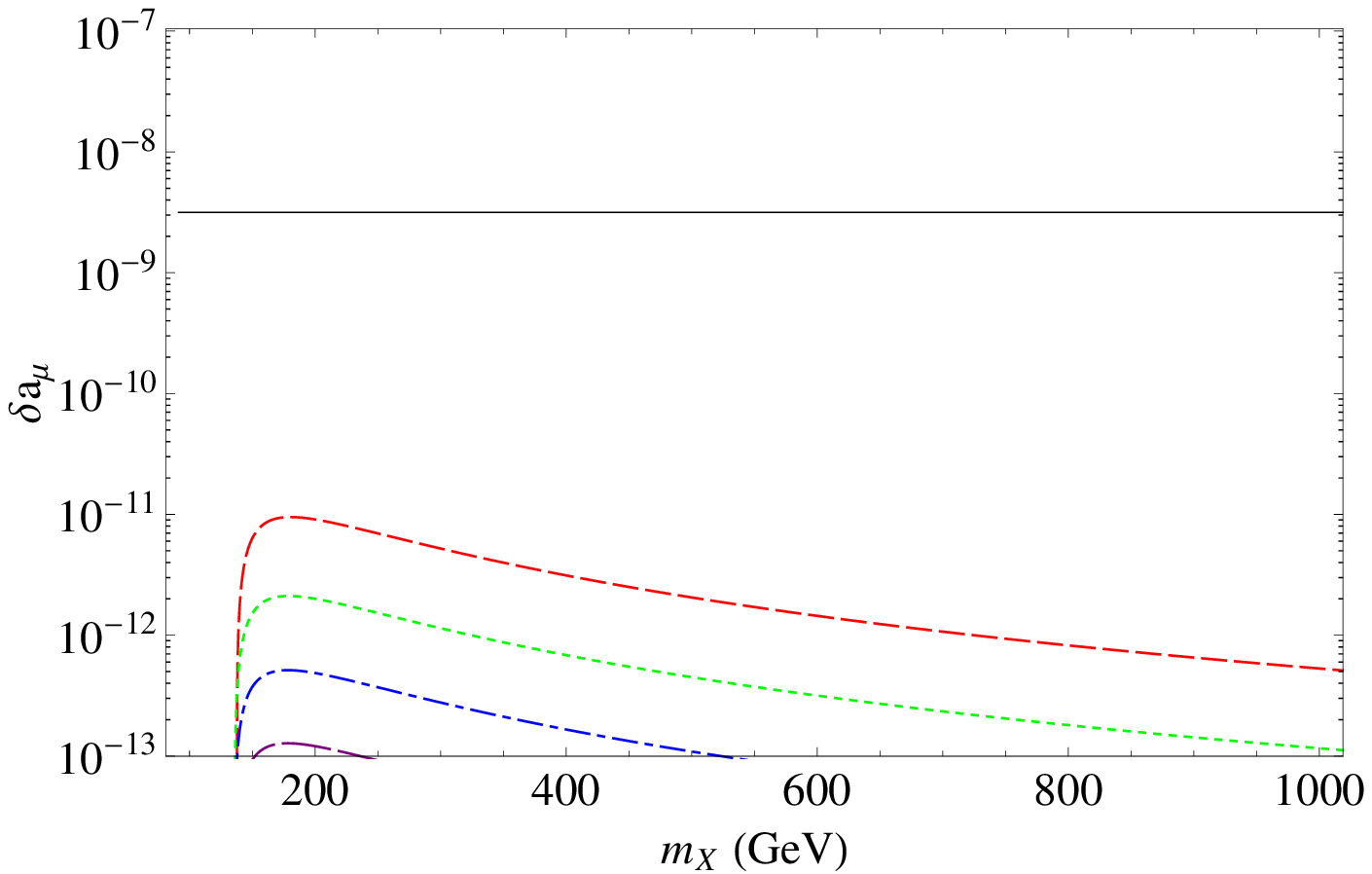}
\end{center}
\caption{Contribution to the anomalous magnetic moment of the muon,
  $\delta a_\mu$ from the hidden $U(1)_X$ model. In the upper panel,
  the red dashed, green dotted, blue dot-dashed, and purple
  long-dashed lines show the four cases $\sin \epsilon =0.03, 0.01,
  0.003,$ and 0.001 for $m_X < m_Z$; in the lower panel,
  the same line--styles and colors show $\sin \epsilon =0.4, 0.2,
  0.1,$ and 0.05 for $m_X > m_Z$.  The horizontal solid line is the
  current limit on the difference between the SM prediction and the
  latest experimental value~\cite{g-2}.  }
\label{Muon_g2}
\end{figure}
%%%%%%%%%%%%%%%%%%%%%%%%%%%%%%%%%%%%%%%%%%%%%%%%%%%%%%%%%%%%%%%%
%
%
The difference between the improved Standard Model (SM) prediction 
of $a_\mu$ and
the latest experimental value for $a_\mu$ is~\cite{g-2}
\begin{equation}
\delta a_\mu = (31.6 \pm 7.9) \cdot 10^{-10}.
\end{equation}
In Fig.~\ref{Muon_g2}, we present the contribution to $a_\mu$ from the
$X$ gauge boson and the modified $Z$ couplings as a function of $m_X$
for various values of $\sin \epsilon$ in the ranges $m_X < m_Z$
(upper panel) and $m_X > m_Z$ (lower panel). As shown in
Fig.~\ref{Muon_g2}, the contribution from our model is well below the
current limit with the exception of very small $m_X$ and sizable $\sin
\epsilon$.

\subsection{Parity-violation effect in atomic physics}

The strength of the vector part of the $Z$ weak neutral current
(i.e. the weak force) between interacting quarks and leptons can be
characterized by their weak charge. This weak charge governs
parity-violation effect in atomic physics. The deviation of the
present experimental results on the weak charge for cesium from the
theoretical SM predictions corresponds to an uncertainty of less than
$1\%$. Consequently, the parity-violation effect in atomic physics can
provide strong constraints for low $m_X$~\cite{Bouchiat04, Fayet07},
if the new effect from the couplings of the $Z$ and $X$ gauge bosons
to electrons and quarks violates parity. Adopting the result in
Ref.~\cite{Fayet07}, the limit on the product of the axial coupling to
the electron and its (average) vector coupling is given by
\begin{equation}
-1.5 \times 10^{-8} {\rm GeV}^{-2} \lesssim {\rm APV} \equiv {g^X_{eA} g^X_{qV}\over m_X^2}
                     + \Delta {g^Z_{eA} g^Z_{qV}\over m_Z^2} \lesssim 0.6
\times 10^{-8} {\rm GeV}^{-2} \,,
\end{equation}
where $g^{Z, X}_{fV} = (g^{Z, X}_{fL}+g^{Z, X}_{fR})/2$ and $g^{Z,
  X}_{fA} = (g^{Z, X}_{fL}-g^{Z, X}_{fR})/2$, and $\Delta$ again
  represents the deviation from the Standard Model value.
%
%
%%%%%%%%%%%%%%%%%%%%%%%%%%%%%%%%%%%%%%%%%%%%%%%%%%%%%%%%%%%%%%%
\begin{figure}
\begin{center}
\includegraphics[width=0.8\linewidth]{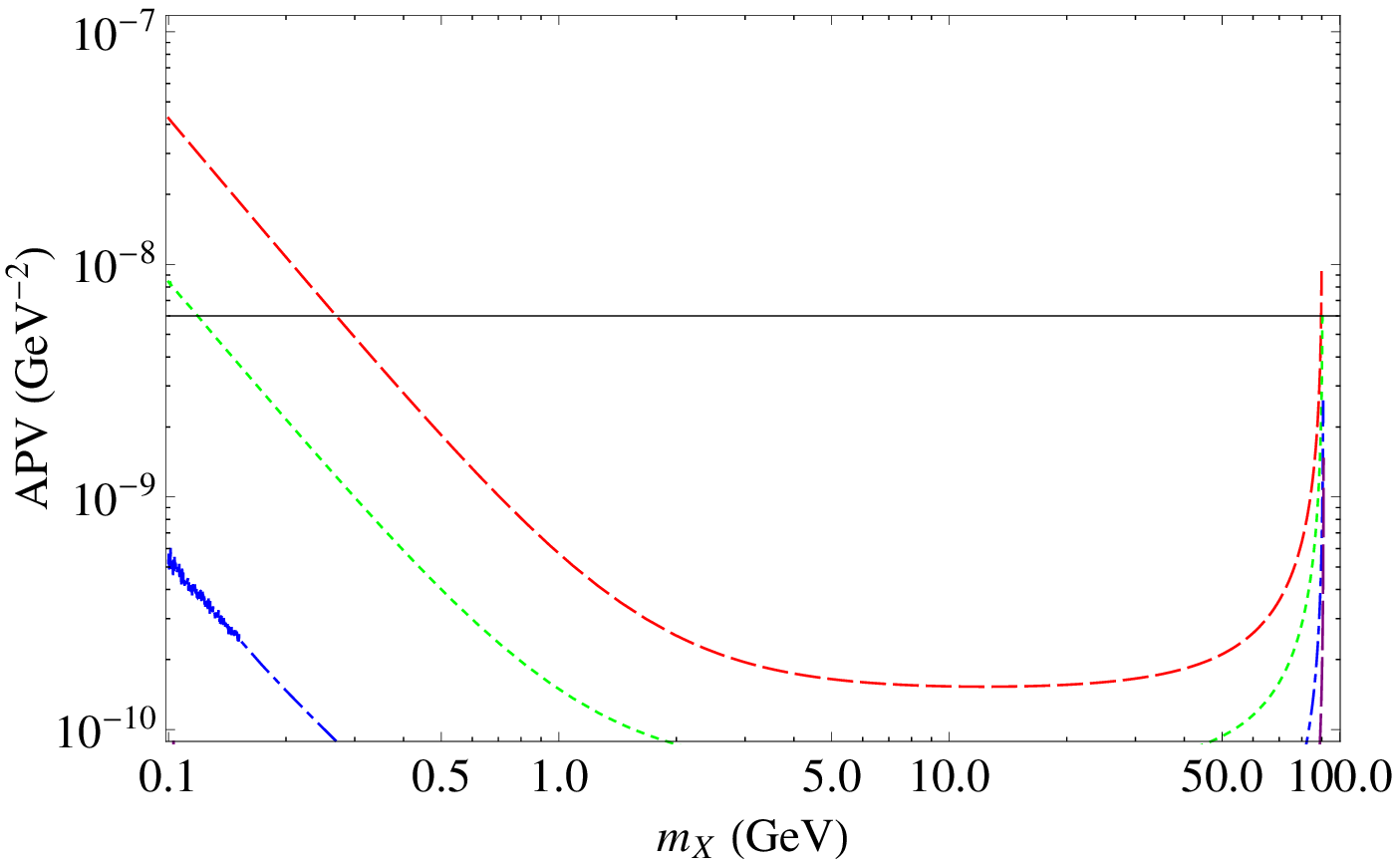}
\includegraphics[width=0.8\linewidth]{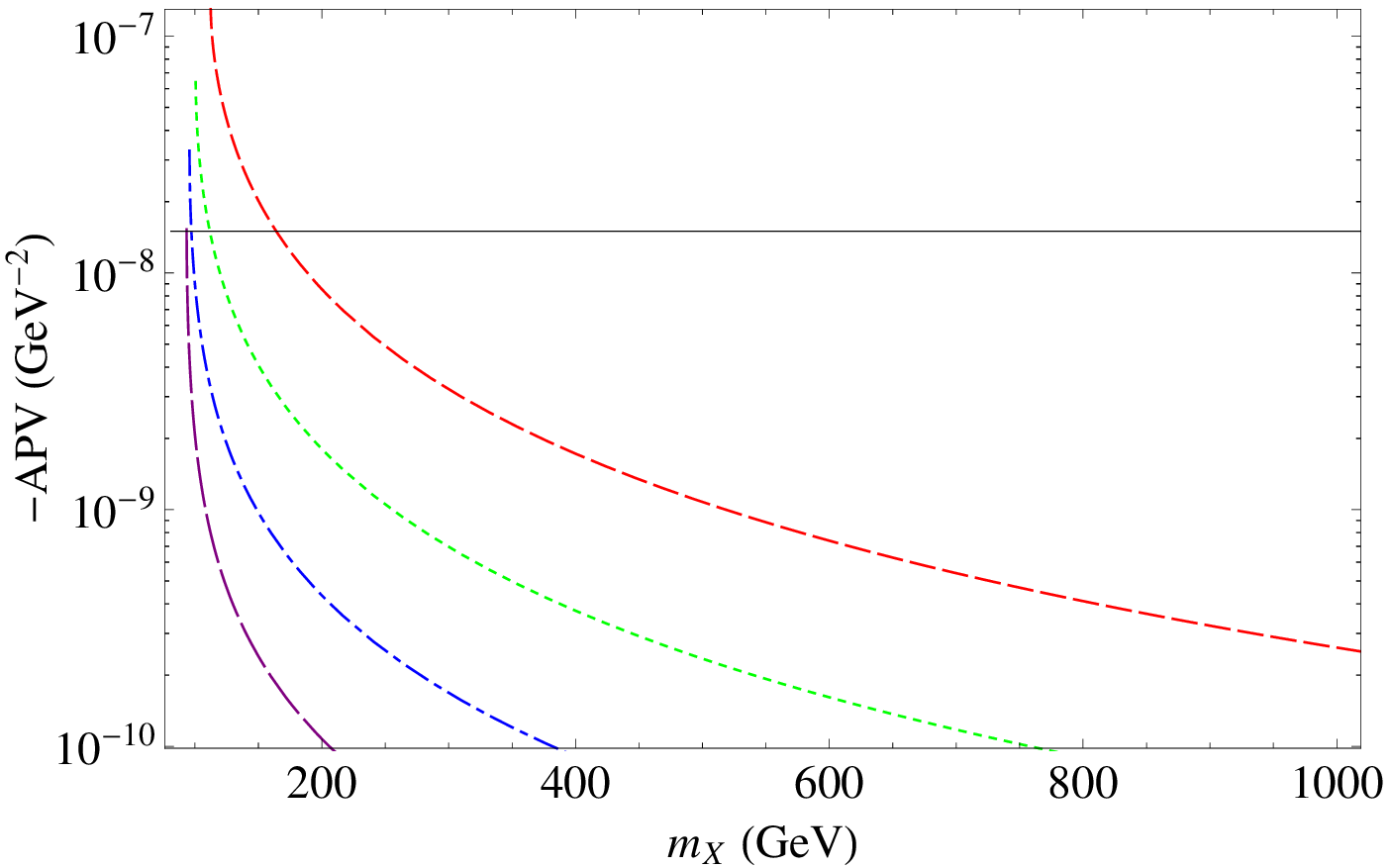}
\end{center}
\caption{Atomic parity-violation (APV) effect from the hidden $U(1)_X$
  model.  In the upper panel, the red dashed, green dotted, blue
  dot-dashed, and purple long-dashed lines show the APV effect corresponding 
  to $\sin \epsilon =0.03, 0.02,
  0.01,$ and 0.005 for $m_X < m_Z$; in the lower panel, the same
  line--styles and colors show $\sin \epsilon =0.4, 0.2, 0.1,$ and
  0.05 for $m_X > m_Z$.  The horizontal solid line represents the
  upper bound given by the difference between the experimental values
  on the weak charge of cesium and the SM
  predictions~\cite{Bouchiat04}.  } \label{APV}
\end{figure}
%%%%%%%%%%%%%%%%%%%%%%%%%%%%%%%%%%%%%%%%%%%%%%%%%%%%%%%%%%%%%%%%
%
%
Fig.~\ref{APV} shows the atomic parity-violation (APV) effect from the
$X$ gauge boson and the modified $Z$ couplings as a function of $m_X$
for $\sin \epsilon =0.03, 0.02, 0.01,$ and 0.005 in the range $m_X <
m_Z$ (upper panel), and $\sin \epsilon =0.4, 0.2, 0.1,$ and 0.05 in
the range $m_X > m_Z$ (lower panel).

%{\bf We should define the
%  symbol APV used in the vertical axis of the figure. Which are the
%  units in the same axis?  Moreover, in the upper panel of the figure
%  the purple curve seems to be missing while the blue one has a
%  strange noisy behavior. Is it OK?}
%Only a small $m_X$ region is
%constrained by the parity-violation effect in atomic physics.

\subsection{$\rho$ parameter}\label{rhoconstraint}

\begin{figure}
\begin{center}
\includegraphics[width=0.83\linewidth]{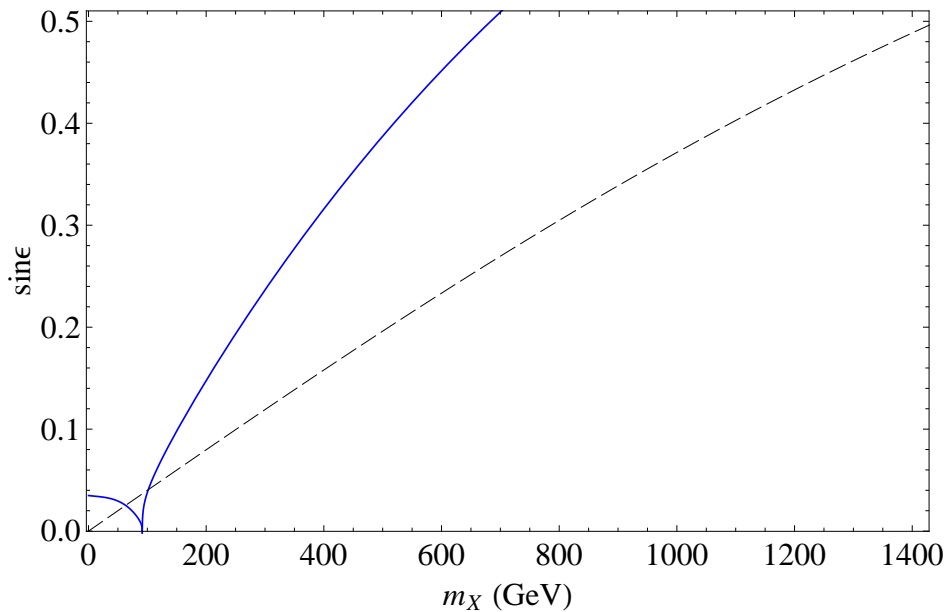}
\end{center}
\caption{Upper bounds on the kinetic mixing parameter $\sin \epsilon$.
The solid line shows the limit from the $\rho$ parameter~\cite{PDG},
while the dashed line shows the limit from EWPT~\cite{Kumar06}.}
\label{mx_se}
\end{figure}

As mentioned in Section~\ref{interactions}, the $\rho$ parameter
is defined as:
\begin{equation}
\rho = {m_W^2\over m_Z^2 c_W^2} = {s_W^2 \over s_{\hat{W}}^2 },
\end{equation}
and the deviation of $\rho$ from 1, $\hat{\delta} \equiv \rho -
1$, is determined by $\sin \epsilon$ according to
Eq.~(\ref{delta}). Therefore, the global fit for the $\rho$
parameter, $\rho-1 = 4^{+8}_{-4} \times 10^{-4}$~\cite{PDG},
results in a limit on the parameters $\epsilon$ and $m_X$.
%
%
%%%%%%%%%%%%%%%%%%%%%%%%%%%%%%%%%%%%%%%%%%%%%%%%%%%%%%%%%%%%%%%
\begin{figure}
\begin{center}
\includegraphics[width=0.8\linewidth]{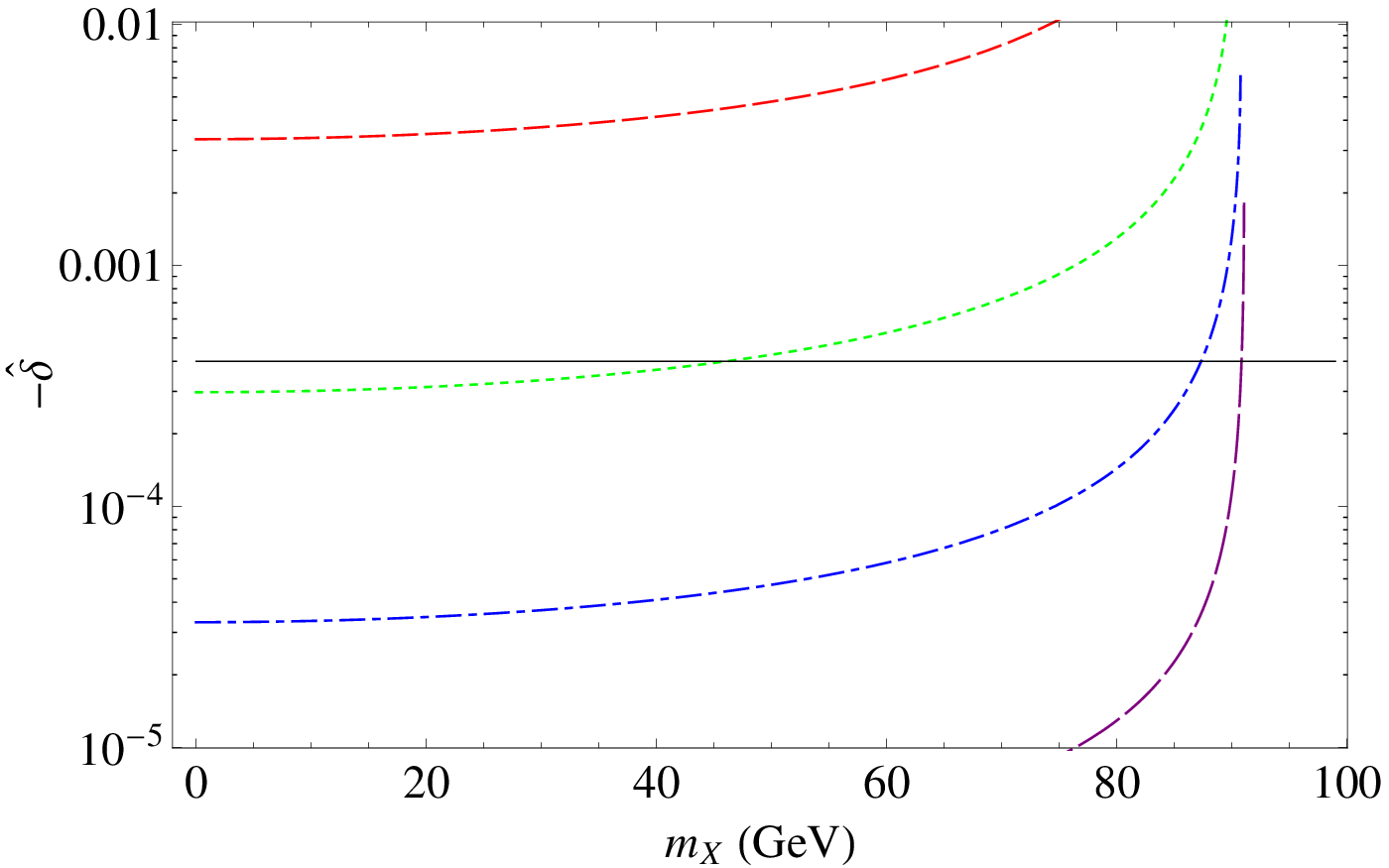}
\includegraphics[width=0.8\linewidth]{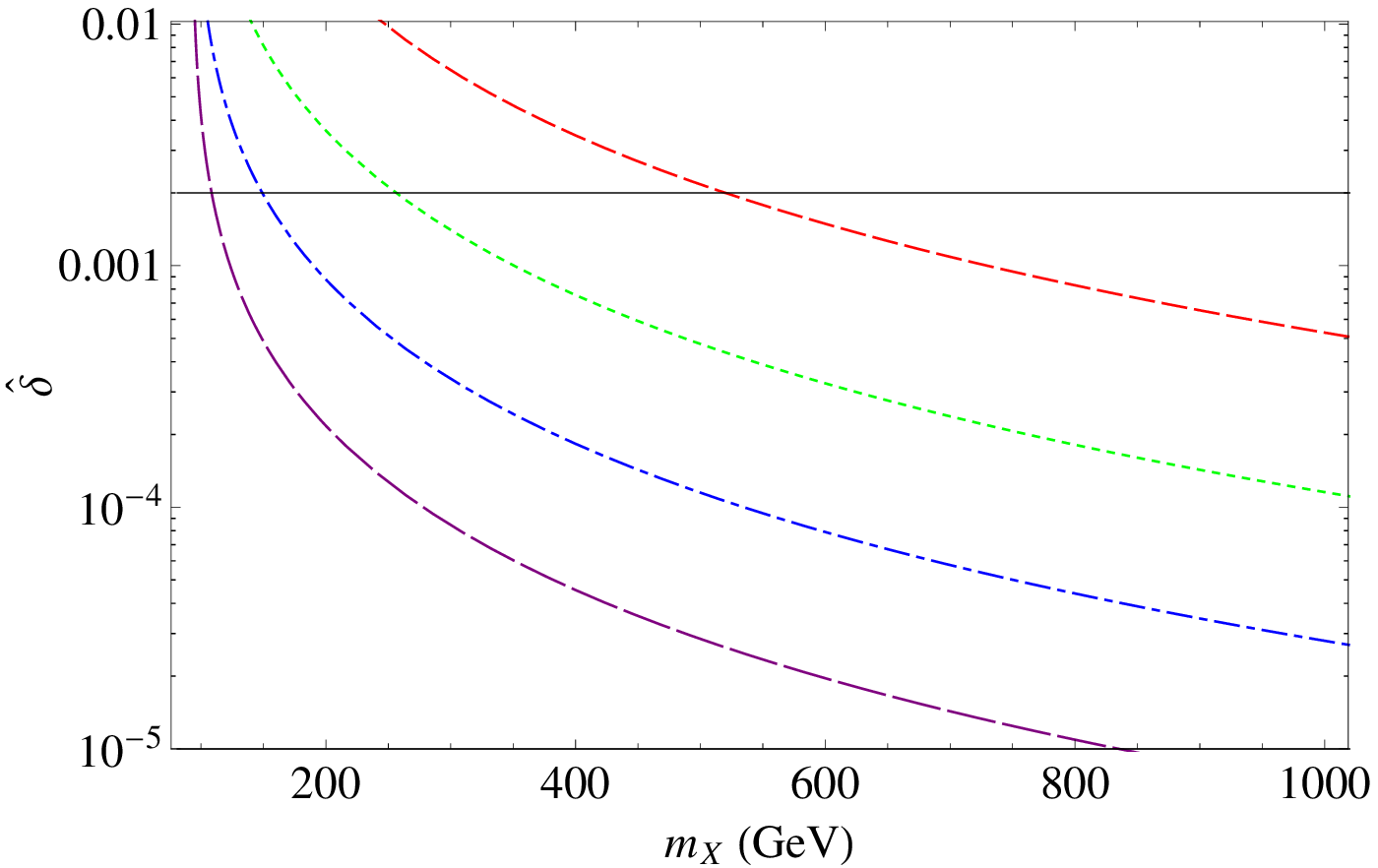}
\end{center}
\caption{Difference from unity of the $\rho$ parameter,
  $\hat{\delta}=\rho-1$, due to the hidden $U(1)_X$ model. In the
  upper panel, the deviation of $\rho$ is shown for $\sin \epsilon
  =0.1, 0.03, 0.01,$ and 0.003 as red dashed, green dotted, blue
  dot-dashed, and purple long-dashed lines, respectively, and for the
  case $m_X < m_Z$. In the lower panel, the same line--styles and
  colors show the cases $\sin \epsilon =0.4, 0.2, 0.1,$ and 0.05 for
  $m_X > m_Z$. The solid horizontal line shows the $2 \sigma$ limit
  from the global fit~\cite{PDG}.}
\label{rho}
\end{figure}
%%%%%%%%%%%%%%%%%%%%%%%%%%%%%%%%%%%%%%%%%%%%%%%%%%%%%%%%%%%%%%%%
%
%
In Fig.~\ref{rho}, we present the deviation of the $\rho$ parameter as
a function of $m_X$ for $\sin \epsilon =0.1, 0.03, 0.01,$ and 0.003 in
the range $m_X < m_Z$ (upper panel) and for $\sin \epsilon =0.4, 0.2,
0.1,$ and 0.05 in the range $m_X > m_Z$ (lower panel). This constraint
is stronger than limits from the $g_\mu-2$ and the APV effect
discussed in the previous Sections. The upper bound on $\sin \epsilon$
from the $\rho$ parameter is shown as a function of $m_X$ in
Fig.~\ref{mx_se}. To obtain this bound we use the $2 \sigma$ limit on
the $\rho$ parameter.

%%%%%%%%%%%%%%%%%%%%%%%%%%%%%%%%%%%%%%%%%%%%%%%%%%%%%%%%%%%%%%
%%%%%%%%%%%%%%%%%%%%%%%%%%%%%%%%%%%%%%%%%%%%%%%%%%%%%%%%%%%%%%%

\subsection{EWPT}
\label{sec:ewpt}

The constraints on the hidden $U(1)_X$ model from electroweak
precision tests (EWPT) have been analyzed in Refs.~\cite{Kumar06,
Chang06}. Here, we adopt the result of Ref.~\cite{Kumar06} which
puts a conservative limit:
\begin{equation}
\left(\tan\epsilon \over 0.1\right)^2 \left(250\mbox{ GeV} \over
m_X \right)^2 \lsim 1 \,.
\end{equation}
This limit is more stringent than that from the $\rho$ parameter
except for $m_X$ around $m_Z$. The $\rho$ parameter and EWPT
bounds in the $\sin\epsilon$--$m_X$ plane are shown in
Fig.~\ref{mx_se}.

\section{Thermal relic abundance of dark matter}\label{abundance}

The relic abundance of the DM candidate $\psi$ depends on the
$\psi\overline{\psi}$ annihilation cross section to Standard Model
particles, which proceeds through $s$-channel exchange of $Z$ and $X$
bosons in the zero-velocity limit. In particular, the annihilation
modes include $\psi\overline{\psi} \rightarrow f\overline{f}, W^+W^-,
Xh,$ and $Zh$, which give the following annihilation cross section
times velocity:
\begin{eqnarray}\label{annihilation}
 \langle\sigma_A v\rangle &\simeq& \sum_f {N_f\over 2\pi}\, m_\psi^2 \sqrt{1-{m_f^2\over m_\psi^2}}
      \left[ (G^2_{fL} + G^2_{fR})\left(1-{m_f^2\over 4 m_\psi^2}\right)
      + {3\over2} G_{fL} G_{fR} {m_f^2\over m_\psi^2} \right] \nonumber\\
    &&  + {1\over \pi}\, m_\psi^2\, G_W^2
      \left(1-{m_W^2\over m_\psi^2}\right)^{3/2}
       \left[ {m_\psi^4\over m_W^4} + 5 {m_\psi^2\over m_W^2} + {3\over 4} \right] \nonumber\\
    &&  + {1\over 8\pi}\, G_{Xh}^2
      \sqrt{1-{\overline{m}_Z^2\over m_\psi^2}
      +\left( {\Delta m_{hZ}^2\over 4m_\psi^2} \right)^2}
       \left\{ 1 + {1\over 2} {m_\psi^2\over m_Z^2}
       \left[ 1 - {\Delta m_{hZ}^2\over 2m_\psi^2}
       +\left( {\Delta m_{hZ}^2\over 4m_\psi^2} \right)^2 \right]  \right\} \nonumber\\
    &&  + {1\over 8\pi}\, G_{Zh}^2
      \sqrt{1-{\overline{m}_X^2\over m_\psi^2}
      +\left( {\Delta m_{hX}^2\over 4m_\psi^2} \right)^2}
       \left\{ 1 + {1\over 2} {m_\psi^2\over m_X^2}
       \left[ 1 - {\Delta m_{hX}^2\over 2m_\psi^2}
       +\left( {\Delta m_{hX}^2\over 4m_\psi^2} \right)^2 \right]  \right\}\,,~~~~~~~
\end{eqnarray}
where:
\begin{eqnarray}
 G_{fL} &=& {g^Z_\psi g^Z_{fL} \over 4m_\psi^2 - m_Z^2} +
            {g^X_\psi g^X_{fL} \over 4m_\psi^2 - m_X^2}\,, \nonumber\\
 G_{fR} &=& {g^Z_\psi g^Z_{fR} \over 4m_\psi^2 - m_Z^2} +
            {g^X_\psi g^X_{fR} \over 4m_\psi^2 - m_X^2}\,, \nonumber\\
 G_W &=& {g_\psi^Z g_W^Z \over  4m_\psi^2-m_Z^2 } +
         {g_\psi^X g_W^X \over  4m_\psi^2-m_X^2 }\,,  \nonumber\\
 G_{Vh} &=& {g_\psi^V g_{XZ}^h \over  4m_\psi^2-m_V^2 } \,.
\label{eq:annihilation_couplings}
\end{eqnarray}
Here, we define $\overline{m}_V^2 = (m_h^2 + m_V^2)/2$ and $\Delta
m_{hV}^2 = m_h^2-m_V^2$.  For the annihilation to $hZ$ ($hX$)
through $s$-channel exchange of the $X$ ($Z$) boson, $(\Delta
m_{hV}^2 / 4m_\psi^2)^2 \ll 1$ is assumed.
% \footnote{In the following analysis, we exclude the contribution of
% some annihilation channels, which are kinematically inaccessible
% depending on the masses $m_\psi, m_X,$ and $m_h$, from
% Eq.~(\ref{annihilation}).}
In the evaluations of the
following Sections, we will assume the numerical value $m_h$=115
GeV.

The present relic density of a hidden Dirac fermion $\psi$ can be
calculated from the following analytic formula:
\begin{equation}\label{Omegah2}
\Omega_{\psi} h^2 \approx\; 2\times \frac{1.07\times10^9\;
{\rm GeV}^{-1}}{M_{pl}} \frac{x_F}{\sqrt{g_*}} \frac{1}
{\langle\sigma_A v\rangle }\;,
\end{equation}
where $g_*$ is the number of relativistic degrees of freedom at the
freeze-out temperature $T_F$ and $x_F \equiv
m_\psi/T_F$~\cite{Bertone04}. The extra factor of two on the
right-hand side of Eq.~(\ref{Omegah2}) results from the fact that the
annihilation can only occur between particle and antiparticle, since
the dark matter candidate is a Dirac fermion.  In the following, we
will use the $\psi$ relic abundance to constrain our parameter space.
In particular, the annihilation cross section $\langle\sigma_A v
\rangle$ given in Eq. (\ref{annihilation}) is dominated either by the
coupling $G_{Xh}$ in the $Xh$ final state channel ($m_X\lsim 120$ GeV)
or by the couplings $G_{fL}$ and $G_{fR}$ in the $f\bar{f}$ final
state.  One can see from Eq.(\ref{eq:interaction_couplings}) that in
both cases $\langle\sigma_A v\rangle$ depends on the parameters $g_X$
and $\epsilon$ through the same multiplicative factor, $g_X
s_{\xi}/c_{\epsilon}$. For each dark matter mass $m_\psi$ and hidden
gauge boson mass $m_X$, we will, therefore, determine the combination
$g_X s_{\xi}/c_{\epsilon}$ by imposing the recent bound on the dark
matter relic density, $(\Omega_{\rm DM} h^2)_{obs} \simeq 0.1123$
\cite{WMAP7}.  Moreover, by fixing $\epsilon$ to its experimental
upper bound, the same combination will be used in Section
\ref{sec:tevatron} to determine the remaining coupling $g_X$.

\section{Direct detection of dark matter}
\label{sec:direct}

Satisfying all the previous constraints, we now discuss prospect of observing DM
in direct detection experiments.
The dark matter particle $\psi$ can elastically scatter off a nucleus
through $t$-channel $X$ and $Z$ gauge boson exchange. The
spin-independent (SI) DM-nucleon scattering cross section can be calculated from the
following effective operator:
\begin{eqnarray}
&& {\cal L}_{eff} = b_f\, \bar{\psi}\gamma_\mu \psi\, \bar{f} \gamma^\mu f\,, \\
\mbox{where} &&
 b_f = {g^Z_\psi (g^Z_{fL}+g^Z_{fR}) \over 2 m_Z^2}
     + {g^X_\psi (g^X_{fL}+g^X_{fR}) \over 2 m_X^2} \,.
\label{eq:bf}
\end{eqnarray}
From the interaction couplings presented in
Section~\ref{interactions}, one calculates $b_u$ and $b_d$ to get
\begin{eqnarray}
 b_p &=& +{\hat{g} g_X\over 4 c_{\hat{W}} } {c_\xi^2\over c_\epsilon} {t_\xi\over m_Z^2}
         \left[ (1-4s_{\hat{W}}^2)\left(1 - {1\over r_X}\right)
         -3 s_{\hat{W}} {t_\epsilon\over t_\xi} \left(t_\xi^2 +  {1\over r_X}\right)
         \right]  \nonumber\\
     &\simeq& {e g_X\over 4 c_W s_W } {c_\xi^2\over c_\epsilon} {t_\xi\over m_Z^2}
         \left\{ (1-4 s_W^2) \left( 1 - {1\over r_X} \right)
         - { 3\over r_X } { s_W t_\epsilon \over t_\xi } \right. \nonumber\\
     &&~~~~~~~~~~~~~~~~~~~ \left. - \omega \left[ 3 + \left( 1 - {1\over r_X} \right)
         \left( {1\over 2} + 2 s_W^2 {1+t_W^2\over 1-t_W^2} \right)
         - {1\over r_X} {s_W t_\epsilon\over t_\xi} {3 t_W^2\over 2 - 2 t_W^2}
          \right]
          \right\},
         \nonumber\\
 b_n &=& -{\hat{g} g_X\over 4 c_{\hat{W}} } {c_\xi^2\over c_\epsilon} {t_\xi\over m_Z^2}
         \left[ \left(1 - {1\over r_X}\right)
         + s_{\hat{W}} {t_\epsilon \over t_\xi} \left(t_\xi^2 + {1\over r_X}\right)
         \right] \nonumber\\
     &\simeq& -{e g_X\over 4 c_W s_W } {c_\xi^2\over c_\epsilon} {t_\xi\over m_Z^2}
         \left\{ \left[ 1 - {1\over r_X} \left(1 - s_W {t_\epsilon\over t_\xi}\right) \right]
         + {\omega\over 2} \left[ 1 + {1\over r_X}
         \left( 1 + {s_W t_W^2 t_\epsilon\over (1-t_W^2) t_\xi } \right) \right] \right\}
         \,.~~~~~~~
\end{eqnarray}
For a given nucleus $^A _Z N$, one has $b_N=Z b_p + (A-Z) b_n$.
Finally, the $\psi$-nucleon elastic scattering cross section is
given by~\cite{Jungman95}:
\begin{equation}
 \sigma_{n,p} = {1\over 64\pi} {\mu_{n,p}^2\over A^2}\, b_N^2\;,
\end{equation}
where $\mu_{n,p}$ is the DM--nucleon reduced mass.
%
%
%%%%%%%%%%%%%%%%%%%%%%%%%%%%%%%%%%%%%%%%%%%%%%%%%%%%%%%%%%%%%%%
\begin{figure}
\begin{center}
\includegraphics[width=0.83\linewidth]{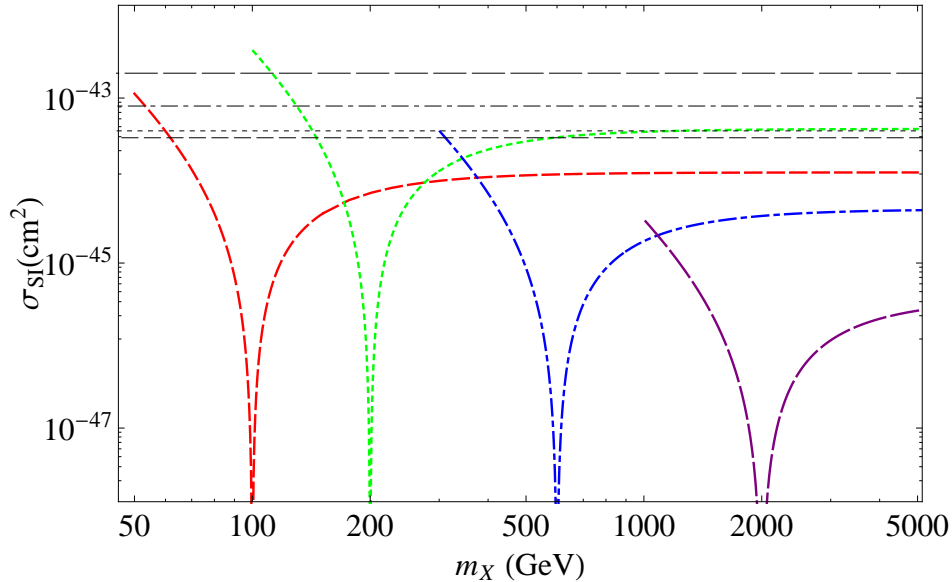}
\end{center}
\caption{Spin-independent DM-nucleon scattering cross section
  $\sigma_{\rm SI}$ calculated by normalizing the DM--nucleon
  couplings to the values that provide the observed DM relic density
  (see text). The red dashed, green dotted, blue dot-dashed, and
  purple long-dashed lines correspond to $m_\psi = 50, 100, 300,$ and
  1000 GeV, respectively. The experimental limits from CDMS
  II~\cite{CDMS2} corresponding to $m_\psi = 100, 300,$ and 1000 GeV
  are shown by the dotted, dot-dashed, and long-dashed lines,
  respectively .  The experimental limit for $m_\psi = 50$, which is
  taken from XENON100~\cite{XENON100}, is shown by the dashed line. }
\label{DirectDetec}
\end{figure}
%%%%%%%%%%%%%%%%%%%%%%%%%%%%%%%%%%%%%%%%%%%%%%%%%%%%%%%%%%%%%%%%
%
%
In Fig.~\ref{DirectDetec}, we present the SI DM-nucleon scattering
cross section as a function of $m_X$ for $m_\psi=50, 100, 300,$ and
1000 GeV. From this figure, one can see that small DM masses around
$m_\psi \sim 100$ GeV are at the level of the current sensitivity of
direct detection experiments. On the other hand, for a larger mass up
to around $m_\psi \sim 300$ GeV and smaller $m_X$, DM signals can be
detectable in the near future.

To calculate the SI DM-nucleon scattering cross sections shown in
Fig.~\ref{DirectDetec}, we have normalized the $b_f$ coefficients given
in Eq. (\ref{eq:bf}) by using the relic abundance constraint. Notice
that, as in the case of the annihilation cross section, also $b_f$
depends on $g_X$ and $\epsilon$ through the multiplicative factor $g_X
s_{\xi}/c_{\epsilon}$. An important consequence of this is
that, at fixed values of $m_X$ and $m_{\psi}$,  the dependence on $g_X$ and
$\epsilon$ cancels out in the ratio
$\sigma_{n,p}/\langle\sigma_A v \rangle$.
 This implies that in our model the direct
detection cross section $\sigma_{n,p}$ is potentially able to put
robust constraints on $m_X$ and $m_{\psi}$, once the $\psi$ particle
is assumed to provide the observed DM relic density in the Universe. On the other
hand, this degeneracy in $g_X$ and $\epsilon$ is not present in the
calculation of accelerator signals, as will be discussed in the next
Section.

\section{Tevatron and LHC probes of $\mathbf{U(1)_X}$ }
\label{sec:tevatron}

For the analysis of collider searches we concentrate on the dimuon
signal from production and decay of the hidden gauge boson $X$ at the
Tevatron and LHC. For this, we need to calculate the corresponding
branching ratios, which depend on the kinematically available decay
channels.  The relevant decay rates of the $X$ gauge boson are given
by:
\begin{eqnarray}
\Gamma(X\rightarrow f\bar{f})&=&\frac{N_f}{12\pi}
\sqrt{\frac{m_X^2}{4}-m_f^2}
\left \{[(g^{X}_{fL})^2+(g^{X}_{fR})^2]\left(1-\frac{m_f^2}{m_X^2}\right)
+6 (g^{X}_{fL})^2 (g^{X}_{fR})^2 \frac{m_f^2}{m_X^2}\right \}\;,\nonumber
\end{eqnarray}
\begin{eqnarray}
\Gamma(X\rightarrow \psi\bar{\psi})&=&
\frac{(g_\psi^X)^2}{6\pi}
\sqrt{\frac{m_X^2}{4}-m_{\psi}^2}
\left(1+2\frac{m_{\psi}^2}{m_X^2} \right)\;,\nonumber
\end{eqnarray}
\begin{eqnarray}
\Gamma(X\rightarrow hZ)&=&\frac{(g^h_{XZ})^2}{48\pi m_X^3}
\sqrt{[m_X^2-(m_Z+m_h)^2][m_X^2-(m_Z-m_h)^2]} \nonumber
\end{eqnarray}
\begin{eqnarray}
&& \times\left[
2+\frac{1}{2}\left(\frac{m_X}{m_Z}+\frac{m_Z}{m_X}-\frac{m_h^2}{m_X
m_Z} \right) \right]\;,
\nonumber\\
\Gamma(X\rightarrow W^{+}W^{-})&=&
\frac{(g_W^X)^2}{192  \pi}m_X \left (\frac{m_X}{m_W} \right )^4
\left (1-4\frac{m_W^2}{m_X^2} \right )^{3/2}
\left [1 + 20\frac{m_W^2}{m_X^2} + 12\frac{m_W^4}{m_X^4} \right ]\;.
\label{eq:x_decay_amplitudes}
\end{eqnarray}

%{\bf [The $g_{XZ}^h$ coupling has dimension of mass, so
%  $\Gamma(X\rightarrow hZ)$ has the right dimension]}

\begin{figure}
\begin{center}
\includegraphics[width=0.83 \linewidth, bb=65 230 517 633]{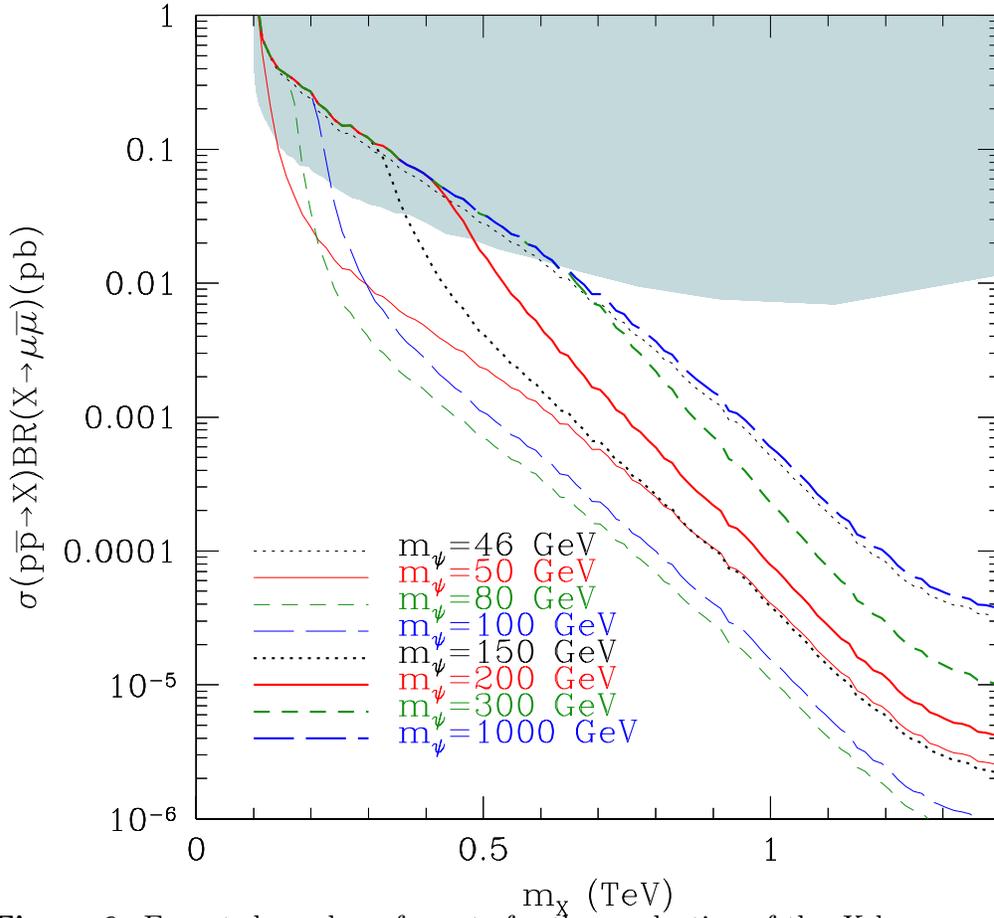}
\end{center}
\caption{Expected number of events for the production of the $X$ boson
and its decay to $\mu\bar{\mu}$ at Tevatron, calculated for
different values of $m_{\psi}$. The shaded area is excluded by dimuon
searches at CDF~\cite{CDF-dimuon}.
Notice that the curve for
  $m_{\psi}$=1000 GeV is cut at $m_X\sim$ 360 GeV due to the
  perturbativity bound shown as a shaded area in
  Fig. \protect\ref{fig:cdf_constraint_mx_mpsi}.
}  \label{fig:sigma_br_cdf}
\end{figure}

We wish now to combine the constraints summarized in the previous
Sections to put bounds on the parameter space of our model, spanned by
the four parameters $m_X$, $m_{\psi}$, $g_X$, and $\sin\epsilon$.  In
particular, the bounds which turn out to be the most constraining, and
that will be the most relevant for the present discussion are: i) the
upper bound on $\sin\epsilon$ from EWPT discussed in Section
\ref{sec:ewpt}; ii) the condition that the $\psi$ relic density
$\Omega_{\psi}h^2$ is equal to the observed dark matter relic density
value $(\Omega_{\rm DM} h^2)_{obs} \simeq 0.1123$; iii) the upper
bound on $\sigma(p\bar{p}\rightarrow X)BR(X\rightarrow \mu\bar{\mu})$
from CDF which will be discussed below in this Section. As a first
approach to this problem, we fix $\sin \epsilon$ to its EWPT upper
bound and $g_X$ to the value required to provide $(\Omega_{\rm DM}
h^2)_{obs}\simeq 0.1123$ and calculate the expected number of dimuon
events at the Tevatron. The result of this exercise is shown in
Fig.~\ref{fig:sigma_br_cdf}, where the curves show the quantity
$\sigma(p\bar{p}\rightarrow X)BR(X\rightarrow\mu\bar{\mu})$ as a
function of $m_X$ at fixed values of $m_{\psi}$. Here and in the
following, we calculate the production cross section
$\sigma(p\bar{p}\rightarrow X)$ by making use of the PYTHIA
code~\cite{pythia}. In the same Figure, the shaded area shows the
upper bound from CDF on the same quantity~\cite{CDF-dimuon}. From this
Figure, one can see that, depending on the value of $m_{\psi}$, the
expected number of dimuon events at the Tevatron
$\sigma(p\bar{p}\rightarrow X)BR(X\rightarrow\mu\bar{\mu})$ can exceed
the CDF upper bound.
%
%
%%%%%%%%%%%%%%%%%%%%%%%%%%%%%%%%%%%%%%%%%%%%%%%%%%%%%%%%%%%%%%%
%%%%%%%%%%%%%%%%%%%%%%%%%%%%%%%%%%%%%%%%%%%%%%%%%%%%%%%%%%%%%%%%
%

\begin{figure}
\begin{center}
\includegraphics[width=0.83 \linewidth, bb=31 220 517 633]{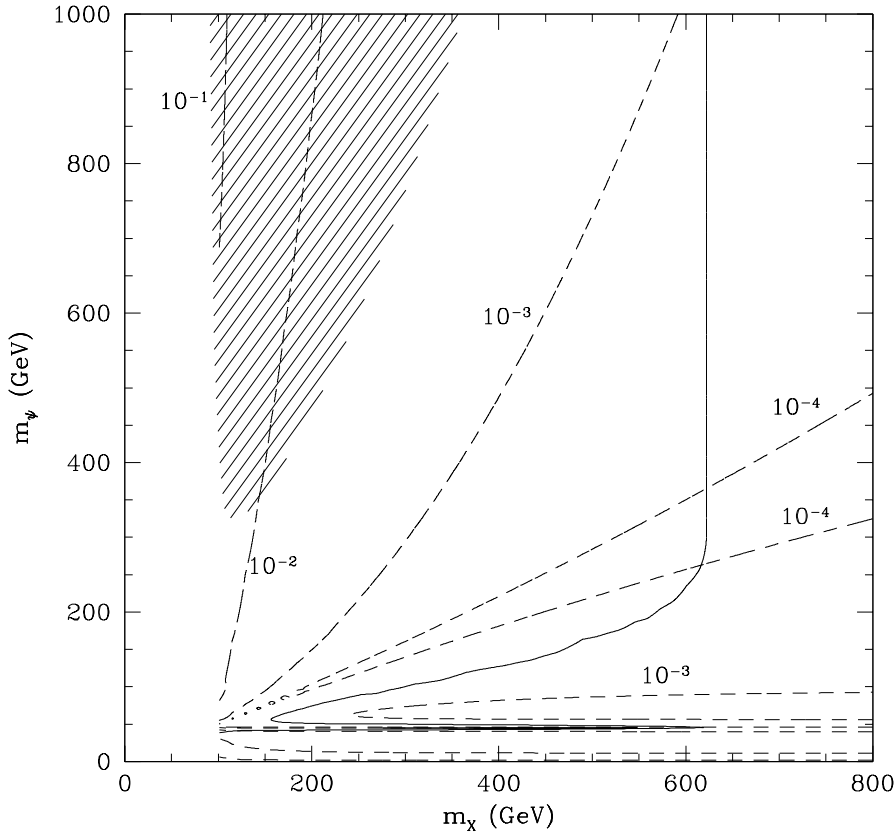}
\end{center}
\caption{The region to the left of the solid line is excluded by CDF
  dimuon $X$ searches in the $m_X$--$m_{\psi}$ plane when the
  $\sin\epsilon$ parameter is set to its upper bound from EWPT.
The dashed lines show the contour plots in the same plane of the minimal
  values of $\sin\epsilon$ compatible to the cosmological constraint
  and with the perturbativity requirement $(g_{\psi}^X)^2/(4\pi)<1$.
Within the shaded area, the minimal value of $\sin\epsilon$ is already exceeding the
upper bound shown in Fig. \protect\ref{fig:ewpt_constraint}
  (see text).}
\label{fig:cdf_constraint_mx_mpsi}
\end{figure}

This is studied in detail in Fig. \ref{fig:cdf_constraint_mx_mpsi}
in the $m_X$--$m_{\psi}$ plane.  In this figure, the region to the
left of the black solid line boundary is excluded by the CDF upper
bound. In particular, one can see from this plot that the range
$m_X\lsim$ 600 GeV is excluded unless $m_{\psi}\lsim$ 200 GeV. In
fact, in this latter case the process $p\bar{p}\rightarrow X
\rightarrow\mu\bar{\mu}$ at the Tevatron can be suppressed
weakening the CDF bound because the invisible decay channel
$X\rightarrow \psi\bar{\psi}$ can become sizable. An exception to
this is when $m_{\psi}\simeq m_Z/2$, because in this case the
resonant annihilation of $\psi$ particles in the calculation of
the relic density requires very low values of the coupling $g_X$
in order to provide the correct amount dark matter. As a
consequence of this, one has $BR(X\rightarrow
\psi\bar{\psi})\simeq 0$ and the constraint on $m_X$ jumps to the
same value $m_X\lsim$ 600 GeV that one finds for higher values of
$m_{\psi}$.
%
%%%%%%%%%%%%%%%%%%%%%%%%%%%%%%%%%%%%%%%%%%%%%%%%%%%%%%%%%%%%%%%
%%%%%%%%%%%%%%%%%%%%%%%%%%%%%%%%%%%%%%%%%%%%%%%%%%%%%%%%%%%%%%%%
%
%
However, in Fig.~\ref{fig:cdf_constraint_mx_mpsi}, the values of $m_X$
and $m_{\psi}$ to the left of the solid line boundary are only
excluded if $\sin\epsilon$ is fixed to its EWPT upper bound. Consequently, in
this region of the $m_X$--$m_{\psi}$ parameter space, the quantity
$\sigma(p\bar{p}\rightarrow X)BR(X\rightarrow\mu\bar{\mu})$ can be
brought below the CDF constraint by using a value for $\sin\epsilon$
smaller than the EWPT bound. In other words,
Fig.~\ref{fig:cdf_constraint_mx_mpsi} shows that for $m_X\lsim$ 600
GeV the CDF constraint can be stronger than the EWPT bound.

\begin{figure}
\begin{center}
  \includegraphics[width=0.83 \linewidth, bb=65 230 517 633]{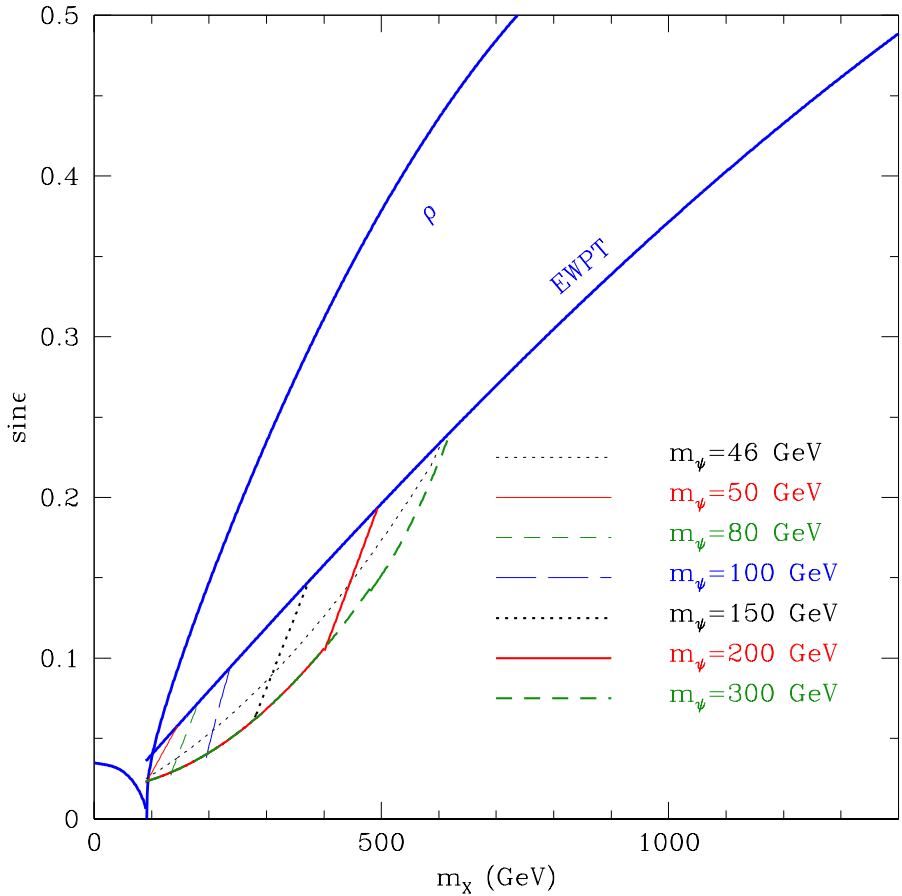}
\end{center}
\caption{Upper bounds on the $\sin\epsilon$ parameter as a function of
  $m_X$ for different values of $m_{\psi}$. The upper thick solid line
  shows the constraint from the $\rho$ parameter~\cite{PDG}, as
  discussed in Section \protect\ref{rhoconstraint}, while the lower
  thick solid line shows the constraint from EWPT~\cite{Kumar06}, as
  discussed in Section \protect\ref{sec:ewpt}. For $m_X\lsim$ 600 GeV,
  the $\sin\epsilon$ parameter is also constrained by dimuon searches
  at CDF~\cite{CDF-dimuon}, depending on the value of $m_{\psi}$, as
  explained in detail in Section
  \protect\ref{sec:tevatron}.} \label{fig:ewpt_constraint}
\end{figure}

In order to see the combined effect of the two constraints on the
parameter space of our model, we provide
Fig.~\ref{fig:ewpt_constraint} where we show the upper bound on
$\sin\epsilon$ as a function of $m_X$ for different fixed values of
$m_{\psi}$ when the CDF constraint is combined to the EWPT one. In
this Figure, the upper thick solid line shows the constraint from the
$\rho$ parameter~\cite{PDG}, as discussed in Section
\protect\ref{rhoconstraint}, while the lower thick solid line shows
the constraint from EWPT~\cite{Kumar06}, as discussed in Section
\protect\ref{sec:ewpt}. The latter bound is mostly more constraining
than the former, and is saturated for $m_X\gsim$ 600 GeV. On the other
hand, the set of curves at $m_X\lsim$ 600 GeV show how the CDF
constraint becomes important at lower $X$ masses for different values
of $m_{\psi}$. As already discussed in connection to
Figs.~\ref{fig:sigma_br_cdf} and \ref{fig:cdf_constraint_mx_mpsi}, one
can see from Fig.~\ref{fig:ewpt_constraint} that the CDF bound on the
parameter space of our model can be stronger than the EWPT one for
$m_X\lsim$ 600 GeV, with the exception $m_\psi \simeq m_Z/2$.

To determine the curves in Fig.~\ref{fig:ewpt_constraint}, the quantity
$\sigma(p\bar{p}\rightarrow X)BR(X\rightarrow\mu\bar{\mu})$ has been
calculated by fixing $g_X$ as before using the relic abundance of the
$\psi$ particle. In particular, following this procedure one has to
check that the upper bound on $\sin\epsilon$ is consistent to the
perturbativity bound $(g_{\psi}^Z)^2/(4 \pi)$,$(g_{\psi}^X)^2/(4
\pi)<$1. Indeed, for some intervals of $m_X$ and $m_{\psi}$, the
combination of the relic abundance, the CDF, and EWPT constraints
require $(g_{\psi}^X)^2/(4 \pi)>$1. The corresponding excluded region
in the $m_X$--$m_{\psi}$ plane is shown as a shaded area in
Fig. \ref{fig:cdf_constraint_mx_mpsi}.  In particular, in this region
of the parameter space, the $f\bar{f}$--final state contribution to the
annihilation cross section $\langle \sigma_A v\rangle$ (which
dominates at larger masses) is suppressed, while the upper bound on
$\sin\epsilon$ is particularly constraining, driving $g_{\psi}^X$ to
large values in order to keep $\Omega_{\psi} h^2$ equal to the
observed cosmological DM abundance.

\begin{figure}
\begin{center}
\includegraphics[width=0.83 \linewidth, bb=65 230 517 633]{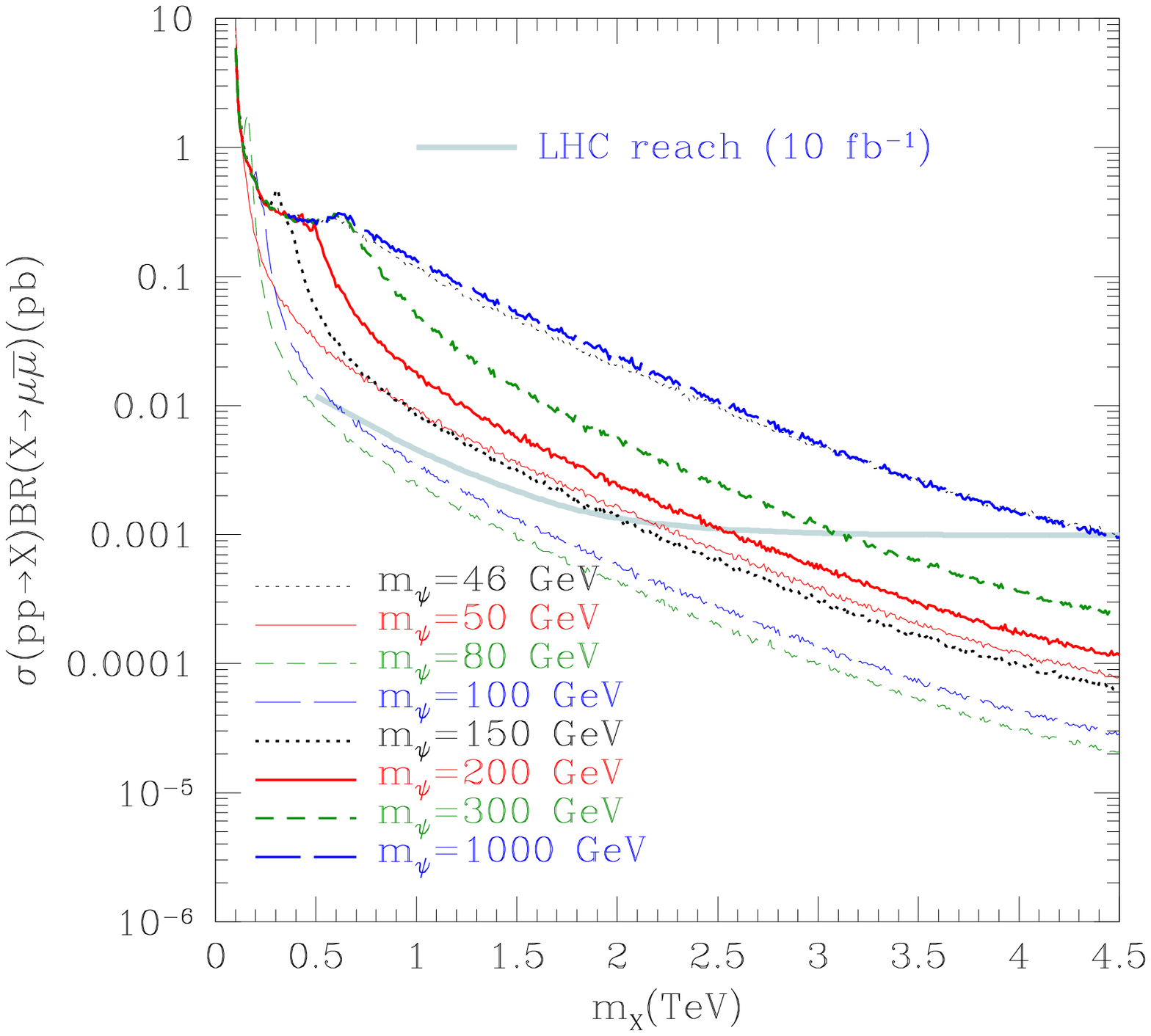}
\end{center}
\caption{Maximal expected number of events for the production of the
  $X$ boson and its decay to $\mu\bar{\mu}$ at LHC for different
  values of $m_{\psi}$ compared to an estimation of the LHC
  sensitivity for an exposure of 10 fb$^{-1}$
  \cite{lhc_discovery_reach}. For each value of $m_{\psi}$, the
  parameter $\epsilon$ is fixed to the corresponding upper bound given
  in Fig. \protect\ref{fig:ewpt_constraint}. Notice that the curve for
  $m_{\psi}$=1000 GeV is cut at $m_X\sim$ 360 GeV due to the
  perturbativity bound shown as a shaded area in
  Fig. \protect\ref{fig:cdf_constraint_mx_mpsi}.} \label{sigma_br_cdf}
\end{figure}

Note that the shaded area excluded by the perturbativity limit does
not extend below $m_X\sim$ 90 GeV. This is due to the fact that in
this region of the parameter space, where the $f\bar{f}$--final state
contribution in $\langle \sigma_A v\rangle$ is suppressed, the
$Xh$-final state takes over as the dominant channel, thanks to the
sizable value of the $t_{\xi}$ parameter in the $g_{XZ}^h$
coupling when $m_X\sim m_Z$ (see Eqs. (\ref{eq:t_csi}) and
(\ref{eq:interaction_couplings})). The ensuing enhancement of $\langle
\sigma_A v\rangle$ implies that for $m_X\lsim$ 90 GeV the observed
relic abundance can be obtained for $(g_{\psi}^Z)^2/(4
\pi)$,$(g_{\psi}^X)^2/(4 \pi)<$1. Notice that, for consistency, the
curves in Figs. \ref{fig:sigma_br_cdf} and \ref{sigma_br_cdf} for
$m_{\psi}$=1000 GeV have been cut at $m_X\sim$ 360 GeV.  Moreover,
since at fixed $m_X$ and $m_{\psi}$, the cosmological constraint fixes
the combination $g_X s_{\xi}/c_{\epsilon}$ which vanishes when
$\epsilon\rightarrow$ 0, keeping the value of $\Omega_{\psi} h^2$
fixed when $\epsilon\rightarrow$ 0 requires $g_X\rightarrow
\infty$. So, imposing $(g_{\psi}^Z)^2/(4 \pi)$,$(g_{\psi}^X)^2/(4
\pi)<$1 allows also to get a lower bound on $\sin\epsilon$ as a
function of $m_X$ and $m_{\psi}$. The result of this procedure is also
shown in Fig.\ref{fig:cdf_constraint_mx_mpsi}, where contour plots at
fixed values of $(\sin\epsilon)_{min}$ are plotted as dashed lines in
the $m_X$--$m_{\psi}$ plane.

%
%%%%%%%%%%%%%%%%%%%%%%%%%%%%%%%%%%%%%%%%%%%%%%%%%%%%%%%%%%%%%%%
%%%%%%%%%%%%%%%%%%%%%%%%%%%%%%%%%%%%%%%%%%%%%%%%%%%%%%%%%%%%%%%%
%
%
The curves given in Fig.~\ref{fig:ewpt_constraint} can be used to
estimate the perspectives of detection of our model at the LHC.  This
is shown in Fig. \ref{sigma_br_cdf}, where the quantity
$\sigma(pp\rightarrow X)BR(X\rightarrow\mu\bar{\mu})$ at the
center-of-mass energy of 14 TeV is shown by a set of lines as a
function of $m_X$ at fixed values of $m_{\psi}$. In each of these
curves, the value of $\sin\epsilon$ is fixed to the upper bound shown
in Fig.  \ref{fig:ewpt_constraint} for the corresponding $m_{\psi}$.
The peculiar hollow shape of some of the curves for $m_X\lsim$ 600 GeV
corresponds to the region of the parameter space where the CDF
constraint on $\sin\epsilon$ overcomes that from EWPT. In the same
Figure, an estimation is also given of the 5 $\sigma$ discovery reach
at the LHC for an exposure of 10 $fb^{-1}$
\cite{lhc_discovery_reach}. We finally recall once again that, at
variance with the analysis of signals at accelerators, the calculation
of the $\psi$-nucleon elastic cross section relevant for direct DM
searches does not rely on any assumption on $\epsilon$ because at
fixed values of $m_X$ and $m_{\psi}$ the cross section $\sigma_{n,p}$
is unambiguously determined by the relic abundance.

%%%%%%%%%%%%%%%%%%%%%%%%%%%%%%%%%%%%%%%%%%%%%%%%%%%%%%%%%%%%%%%
%%%%%%%%%%%%%%%%%%%%%%%%%%%%%%%%%%%%%%%%%%%%%%%%%%%%%%%%%%%%%%%%
%
%
%

\section{Conclusions}
\label{sec:conclusions}

We postulated the existence of a hidden $U(1)_X$ gauge symmetry and a
Dirac dark matter fermion charged under $U(1)_X$.  Assuming that the
hidden sector communicates with the Standard Model sector only through
the kinetic mixing between $U(1)_X$ and $U(1)_Y$, the phenomenology of
such scenario depends on four parameters: the kinetic mixing angle,
the mass of the $X$ boson, the dark matter mass, the coupling between
the DM particle and the $X$ gauge boson.  We have considered various
observables constraining the kinetic mixing parameter and the $U(1)_X$
coupling constant as a function of the $X$ gauge boson mass.  Since
the dark matter annihilation to the Standard Model fields proceeds
through kinetic mixing, we have used the relic density of the DM
particle (determined by the standard thermal freeze-out process) to
constrain a combination of the kinetic mixing and the $X$ gauge boson
coupling, requiring $\Omega_{\psi} h^2$ to be equal to the observed DM
density in the Universe. Saturating all these constraints, we analyzed
the spin-independent elastic cross-section of the dark matter off
nucleons showing that a large parameter space is within the
sensitivity of future direct detection experiments.  We have also
analyzed collider searches of the $X$ gauge boson, concentrating on
the dimuon signal at the Tevatron and the LHC.  In particular, we
found that the current Tevatron result puts the stronger bound on the
kinetic mixing term $\epsilon$ for $m_X\lsim$ 600 GeV, while at larger
masses Electro--Weak Precision Tests are more constraining.  We have
also found some intervals in the $m_X$ and $m_{\psi}$ masses
($m_X\lsim$ 350 GeV and $m_{\psi}\gsim$ 320 GeV) that result to be
excluded by a perturbativity requirement, since the combination of the
relic abundance, the CDF and EWPT constraints require
$(g_{\psi}^X)^2/(4 \pi)>$1. The same perturbativity constraint,
combined to the requirement that the relic density of our DM candidate
matches the observed value, allowed us to put also a lower bound on
the $s_{\epsilon}$ parameter as a function of $m_{\chi}$ and
$m_{\psi}$.

Finally, the LHC prospects for observing the $X$ gauge boson have been
analyzed taking the center-of-mass energy of 14 TeV and the integrated
luminosity of 10 fb$^{-1}$.  We found that an $X$ gauge boson mass up
to 4.5 TeV is within the LHC reach when the dark matter mass
$m_{\psi}$ is heavy enough to suppress the invisible decay
$X\rightarrow \psi\bar{\psi}$, or in the resonant limit $m_\psi\sim
m_Z/2$, when the $X\rightarrow \psi\bar{\psi}$ decay is negligible
because the $X$--dark matter coupling $g_X$ is suppressed due to the
cosmological bound on $\Omega_{\psi} h^2$. Moreover, an $X$ boson mass
up to 2.5 TeV can be probed if the dark matter mass is heavier than
about 200 GeV.  These results are obtained assuming that the kinetic
mixing parameter saturates all the constraints including the Tevatron
limit, and would be weakened by choosing smaller values.

\medskip

{\bf Acknowledgment:}
SS is supported by NRF with CQUEST grant 2005-0049049 and by by the
Sogang Research Grant 2010.

\begin{appendix}
\section{Interaction couplings}\label{couplings}
\label{sec:appendix}

Comparing Eq.~(\ref{g's}) with Eqs.~(\ref{interaction-Wff}) -
(\ref{interaction-hVV}), one can easily find the redefined
couplings expressed by the physical observables (unhatted parameters):
\begin{eqnarray}
g_f^W &=& - {e\over \sqrt{2} s_W}
 \left(1-{\omega \over 2(1-t_W^2)}\right)\,, \nonumber \\
g^Z_{fL} &=& -{e\over c_{{W}} s_{{W}} }\, c_\xi \,
         \left\{ T_3 \left[1+ {\omega \over 2}\right]
         - Q \left[s_{{W}}^2 + \omega \left( {2 - t_W^2
         \over 2( 1- t_W^2) }\right) \right] \right\}\,, \nonumber\\
g^Z_{fR} &=& {e\over c_{{W}} s_{{W}} }\, c_\xi \,
         Q \left[s_{{W}}^2 + \omega \left( {2 - t_W^2
         \over 2( 1- t_W^2) }\right) \right]\,,    \nonumber\\
g^Z_\psi &=& - g_X {s_\xi\over c_\epsilon}\,, \nonumber\\
g^X_{fL} &=& - {e\over c_{{W}} s_{{W}} }\, {c_\xi} \left\{
         T_3 \left[ s_W t_\epsilon - t_\xi + {1\over 2}\,
         \omega \left(t_\xi + { s_W t_W^2 t_\epsilon\over 1 - t_W^2} \right) \right]
         \right. \nonumber\\
  && ~~~~~~~~~~~~~~\left. + Q \left[ s_W^2 t_\xi - s_W t_\epsilon
         + {1\over 2}\, t_W^2 \omega \left({t_\xi -s_W t_\epsilon \over 1-t_W^2 } \right) \right]
         \right\}\,, \nonumber\\
g^X_{fR} &=& -{e\over c_{{W}} s_{{W}} }\, {c_\xi}\, Q \left[ s_W^2 t_\xi - s_W t_\epsilon
         + {1\over 2}\, t_W^2 \omega \left({t_\xi -s_W t_\epsilon \over 1-t_W^2 } \right)
         \right]\,,  \nonumber\\
g^X_\psi &=& - g_X {c_\xi\over c_\epsilon}\,, \nonumber
\end{eqnarray}

\begin{eqnarray}
g_W^Z &=& {e \over t_{{W}}}\, c_\xi \left(1-{\omega \over 2( c_W^2-s_W^2)} \right)\,, \nonumber\\
g_W^X &=& - {e \over t_{{W}}}\, s_\xi \left(1-{\omega \over 2( c_W^2-s_W^2)} \right)\,, \nonumber\\
g^h_{ZZ} &=& {m_{{Z}}^2\over v}\, c_\xi^2\, (1 + \omega)\,, \nonumber\\
g^h_{XX} &=& {m_{{Z}}^2\over v}\, c_\xi^2\, \left[ t_\xi^2 + s_W^2 t_\epsilon^2
               - \omega \left(2 + t_\xi^2 - { s_W^2 t_W^2 t_\epsilon^2 \over
               1 - t_W^2} \right) \right]\,, \nonumber\\
g^h_{XZ} &=& {m_{{Z}}^2\over v}\, c_\xi^2\, 2 \left[ 2s_W t_\epsilon
               - t_\xi + \omega \left( t_\xi + {s_W t_W^2 t_\epsilon
               \over 1- t_W^2} \right) \right]\,.
\label{eq:interaction_couplings}
\end{eqnarray}

\end{appendix}


\begin{thebibliography}{99}

\bibitem{Langacker08}
%\cite{Langacker:2008yv}
%\bibitem{Langacker:2008yv}
  For a review, see, P.~Langacker,
  %``The Physics of Heavy $Z^\prime$ Gauge Bosons,''
  Rev.\ Mod.\ Phys.\  {\bf 81} (2008) 1199
  [arXiv:0801.1345 [hep-ph]].
  %%CITATION = RMPHA,81,1199;%%


\bibitem{Holdom85}
  B.~Holdom,
  %``Two U(1)'S And Epsilon Charge Shifts,''
  Phys.\ Lett.\  B {\bf 166}, 196 (1986).


\bibitem{Huh07}
  J.~H.~Huh, J.~E.~Kim, J.~C.~Park and S.~C.~Park,
  %``Galactic 511 keV line from MeV milli-charged dark matter,''
  Phys.\ Rev.\  D {\bf 77}, 123503 (2008)
  [arXiv:0711.3528 [astro-ph]].


  \bibitem{Chun08}
%\cite{Chun:2008by}
%\bibitem{Chun:2008by}
  E.~J.~Chun and J.~C.~Park,
  %``Dark matter and sub-GeV hidden U(1) in GMSB models,''
  JCAP {\bf 0902} (2009) 026
  [arXiv:0812.0308 [hep-ph]].
  %%CITATION = JCAPA,0902,026;%%

\bibitem{Kang10}
  Z.~Kang, T.~Li, T.~Liu, C.~Tong and J.~M.~Yang,
  %``Light Dark Matter from the U(1)_X Sector in the NMSSM with Gauge
  %Mediation,''
  arXiv:1008.5243 [hep-ph].
  %%CITATION = ARXIV:1008.5243;%%


\bibitem{Gopalakrishna09}
%\cite{Gopalakrishna:2009yz}
%\bibitem{Gopalakrishna:2009yz}
  S.~Gopalakrishna, S.~J.~Lee and J.~D.~Wells,
  %``Dark matter and Higgs boson collider implications of fermions in an
  %abelian-gauged hidden sector,''
  Phys.\ Lett.\  B {\bf 680} (2009) 88
  [arXiv:0904.2007 [hep-ph]].
  %%CITATION = PHLTA,B680,88;%%

\bibitem{Mambrini10}
%\cite{Mambrini:2010dq}
%\bibitem{Mambrini:2010dq}
  Y.~Mambrini,
  %``The kinetic dark-mixing in the light of CoGENT and XENON100,''
  arXiv:1006.3318 [hep-ph].
  %%CITATION = ARXIV:1006.3318;%%

\bibitem{Cheung07}
  K.~Cheung and T.~C.~Yuan,
  %``Hidden fermion as milli-charged dark matter in Stueckelberg Z' model,''
  JHEP {\bf 0703}, 120 (2007)
  [arXiv:hep-ph/0701107].

\bibitem{Feldman07}
  D.~Feldman, Z.~Liu and P.~Nath,
  %``The Stueckelberg Z' extension with kinetic mixing and milli-charged dark
  %matter from the hidden sector,''
  Phys.\ Rev.\  D {\bf 75}, 115001 (2007)
  [arXiv:hep-ph/0702123].

\bibitem{Fucito08}
  F.~Fucito, A.~Lionetto, A.~Mammarella and A.~Racioppi,
  %``St\'uckelino Dark Matter in Anomalous U(1)' Models,''
  Eur.\ Phys.\ J.\  C {\bf 69}, 455 (2010)
  [arXiv:0811.1953 [hep-ph]].
  %%CITATION = EPHJA,C69,455;%%


\bibitem{Bouchiat04}
  C.~Bouchiat and P.~Fayet,
  %``Constraints on the parity-violating couplings of a new gauge boson,''
  Phys.\ Lett.\  B {\bf 608}, 87 (2005)
  [arXiv:hep-ph/0410260].

\bibitem{Fayet07}
  P.~Fayet,
  %``U-boson production in e+ e- annihilations, psi and Upsilon decays, and
  %light dark matter,''
  Phys.\ Rev.\  D {\bf 75}, 115017 (2007)
 [arXiv:hep-ph/0702176].



\bibitem{Kumar06}
%\cite{Kumar:2006gm}
%\bibitem{Kumar:2006gm}
  J.~Kumar and J.~D.~Wells,
  %``LHC and ILC probes of hidden-sector gauge bosons,''
  Phys.\ Rev.\  D {\bf 74} (2006) 115017
  [arXiv:hep-ph/0606183].
  %%CITATION = PHRVA,D74,115017;%%

\bibitem{Chang06}
%\cite{Chang:2006fp}
%\bibitem{Chang:2006fp}
  W.~F.~Chang, J.~N.~Ng and J.~M.~S.~Wu,
  %``A Very Narrow Shadow Extra Z-boson at Colliders,''
  Phys.\ Rev.\  D {\bf 74} (2006) 095005
  [Erratum-ibid.\  D {\bf 79} (2009) 039902]
  [arXiv:hep-ph/0608068].
  %%CITATION = PHRVA,D74,095005;%%

\bibitem{CDMS2}
  Z.~Ahmed {\it et al.}  [The CDMS-II Collaboration],
  %``Results from the Final Exposure of the CDMS II Experiment,''
  arXiv:0912.3592 [astro-ph.CO].

\bibitem{XENON100}
  E.~Aprile {\it et al.}  [XENON100 Collaboration],
  %``First Dark Matter Results from the XENON100 Experiment,''
  arXiv:1005.0380 [astro-ph.CO].

\bibitem{babu97}
  K.~S.~Babu, C.~F.~Kolda and J.~March-Russell,
  %``Implications of generalized Z Z' mixing,''
  Phys.\ Rev.\  D {\bf 57}, 6788 (1998)
  [arXiv:hep-ph/9710441].

\bibitem{PDG}
  C.~Amsler {\it et al.}  [Particle Data Group],
  %``Review of particle physics,''
  Phys.\ Lett.\  B {\bf 667}, 1 (2008).

\bibitem{g-2}
  T.~Teubner, K.~Hagiwara, R.~Liao, A.~D.~Martin and D.~Nomura,
  %``Update of g-2 of the muon and Delta alpha,''
  arXiv:1001.5401 [hep-ph].

\bibitem{Bertone04}
  G.~Bertone, D.~Hooper and J.~Silk,
  %``Particle dark matter: Evidence, candidates and constraints,''
  Phys.\ Rept.\  {\bf 405}, 279 (2005)
  [arXiv:hep-ph/0404175].

\bibitem{WMAP7}
  E.~Komatsu {\it et al.},
  %``Seven-Year Wilkinson Microwave Anisotropy Probe (WMAP) Observations:
  %Cosmological Interpretation,''
  arXiv:1001.4538 [astro-ph.CO].

\bibitem{Jungman95}
  G.~Jungman, M.~Kamionkowski and K.~Griest,
  %``Supersymmetric dark matter,''
  Phys.\ Rept.\  {\bf 267}, 195 (1996)
  [arXiv:hep-ph/9506380].

\bibitem{pythia} {\tt http://home.thep.lu.se/~torbjorn/Pythia.html}


\bibitem{CDF-dimuon}
  T.~Aaltonen {\it et al.}  [CDF Collaboration],
  %``A search for high-mass resonances decaying to dimuons at CDF,''
  Phys.\ Rev.\ Lett.\  {\bf 102}, 091805 (2009)
  [arXiv:0811.0053 [hep-ex]];
  C. Ciobanu et al., FERMILAB-FN-0773-E (2008).


\bibitem{lhc_discovery_reach}
  I.~Golutvin, P.~Moissenz, V.~Palichik, M.~Savina and S.~Shmatov,
  %``Search for TeV-scale bosons in the dimuon channel with the CMS  detector,''
  Czech.\ J.\ Phys.\  {\bf 54}, A261 (2004)
  [arXiv:hep-ph/0310336].
  %%CITATION = CZYPA,54,A261;%%

%--------------------------------------------------
%
%{\tt References to be arranged properly.}
%
%--------------------------------------------------
%
%\bibitem{Pospelov07}
%  M.~Pospelov, A.~Ritz and M.~B.~Voloshin,
%  %``Secluded WIMP Dark Matter,''
%  Phys.\ Lett.\  B {\bf 662}, 53 (2008)
%  [arXiv:0711.4866 [hep-ph]].
%
%\bibitem{baek08}
%  S.~Baek and P.~Ko,
%  %``Phenomenology of $U(1)_{L_\mu - L_\tau}$ charged dark matter at PAMELA and
%  %colliders,''
%  JCAP {\bf 0910}, 011 (2009)
%  [arXiv:0811.1646 [hep-ph]].
%
%\bibitem{Bennett99}
%  S.~C.~Bennett and C.~E.~Wieman,
%  %``Measurement of the 6S --> 7S transition polarizability in atomic cesium
%  %and an improved test of the standard model,''
%  Phys.\ Rev.\ Lett.\  {\bf 82}, 2484 (1999)
%  [Erratum-ibid.\  {\bf 83}, 889 (1999)]
%  [arXiv:hep-ex/9903022].
%
%\bibitem{Ginges03}
%  J.~S.~M.~Ginges and V.~V.~Flambaum,
%  %``Violations of fundamental symmetries in atoms and tests of unification
%  %theories of elementary particles,''
%  Phys.\ Rept.\  {\bf 397}, 63 (2004)
%  [arXiv:physics/0309054].
%
%\bibitem{salvioni09}
%  E.~Salvioni, G.~Villadoro and F.~Zwirner,
%  %``Minimal Z' models: present bounds and early LHC reach,''
%  JHEP {\bf 0911}, 068 (2009)
%  [arXiv:0909.1320 [hep-ph]].
%
%\bibitem{cacciapaglia06}
%  G.~Cacciapaglia, C.~Csaki, G.~Marandella and A.~Strumia,
%  %``The minimal set of electroweak precision parameters,''
%  Phys.\ Rev.\  D {\bf 74}, 033011 (2006)
%  [arXiv:hep-ph/0604111].
%
%\bibitem{Khalil10}
%  S.~Khalil, H.~S.~Lee and E.~Ma,
%  %``Bound on Z' Mass from CDMS II in the Dark Left-Right Gauge Model II,''
%  Phys.\ Rev.\  D {\bf 81}, 051702 (2010)
%  [arXiv:1002.0692 [hep-ph]].
%
%\bibitem{Cheung10}
%  K.~Cheung, K.~H.~Tsao and T.~C.~Yuan,
%  %``Hidden Sector Dirac Dark Matter, Stueckelberg Z' Model and the CDMS
%  %Experiment,''
%  arXiv:1003.4611 [hep-ph].
%
%\bibitem{Feldman10}
%  D.~Feldman, Z.~Liu, P.~Nath and G.~Peim,
%  %``Multicomponent Dark Matter in Supersymmetric Hidden Sector Extensions,''
%  Phys.\ Rev.\  D {\bf 81}, 095017 (2010)
%  [arXiv:1004.0649 [hep-ph]].
%

\end{thebibliography}
\end{document}